\documentclass{article}

\usepackage{arxiv}
\usepackage[dvipsnames]{xcolor}
\usepackage[utf8]{inputenc} % allow utf-8 input
\usepackage[T1]{fontenc}    % use 8-bit T1 fonts
\usepackage[colorlinks=true]{hyperref}
\hypersetup{
colorlinks=true,
linkcolor={red},
urlcolor={blue},
filecolor={black},
citecolor={Plum},
}% hyperlinks
\usepackage{url}            % simple URL typesetting
\usepackage{booktabs}       % professional-quality tables
\usepackage{amsfonts}       % blackboard math symbols
\usepackage{nicefrac}       % compact symbols for 1/2, etc.
\usepackage{microtype}      % microtypography
\usepackage{graphicx}
\usepackage[square, colon, numbers]{natbib}
\usepackage{doi}
\usepackage[labelfont=bf]{caption}
\usepackage{siunitx}
\usepackage{tikz}
\usepackage{amsmath}

\def\checkmark{\tikz\fill[scale=0.4](0,.35) -- (.25,0) -- (1,.7) -- (.25,.15) -- cycle;}
\def\goldmedal{\textsuperscript{\includegraphics[scale=0.015]{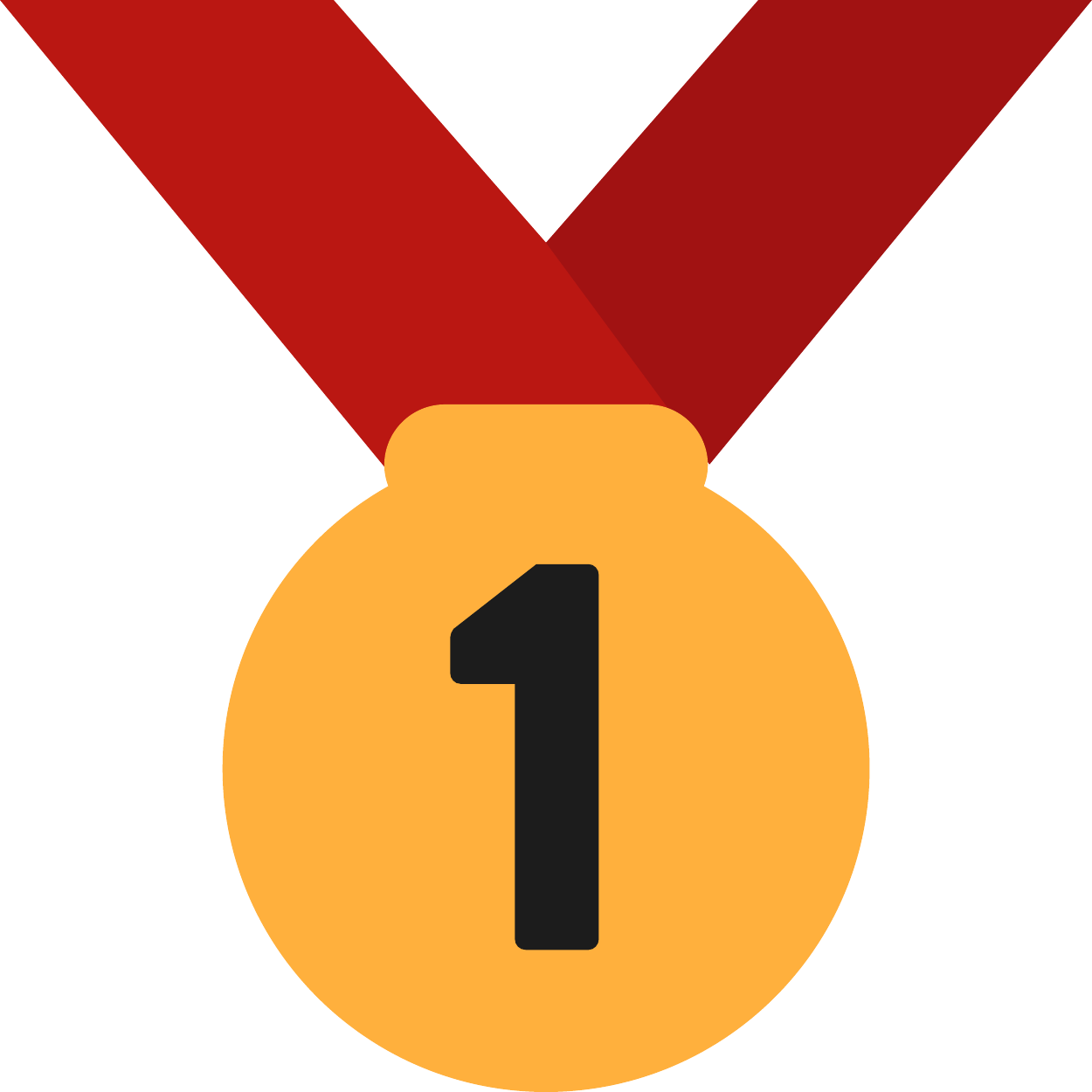}}}
\def\silvermedal{\textsuperscript{\includegraphics[scale=0.015]{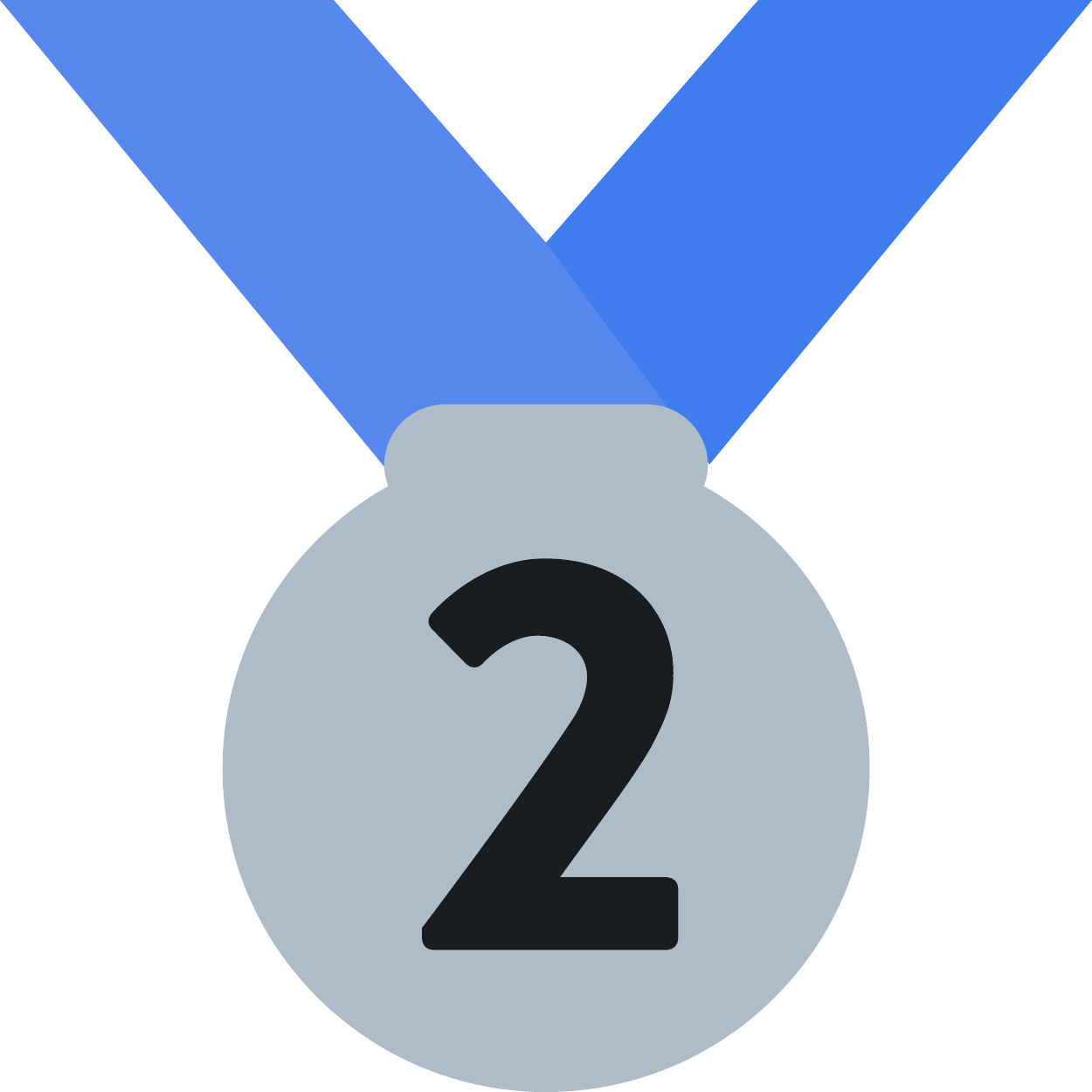}}}
\def\bronzemedal{\textsuperscript{\includegraphics[scale=0.015]{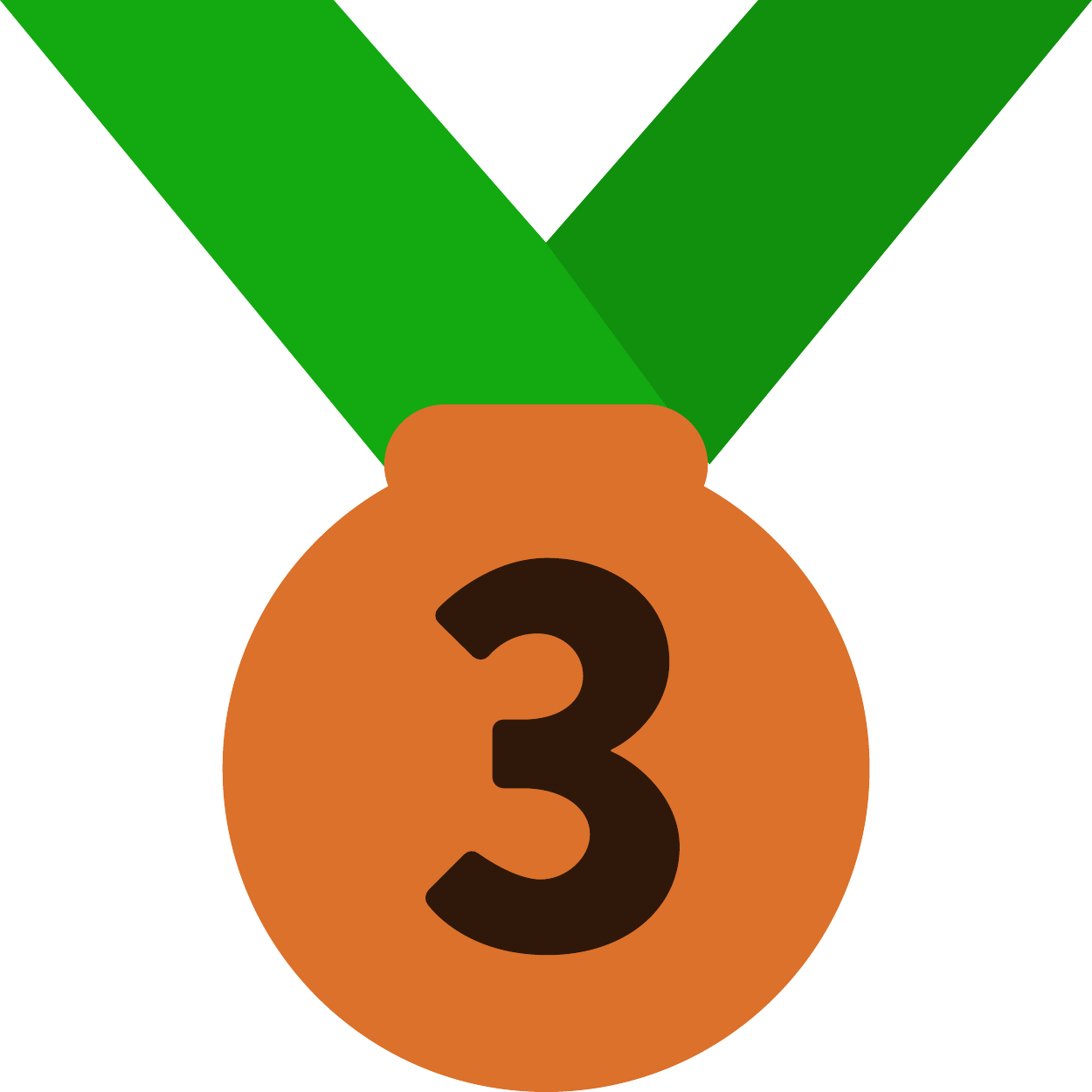}}}

\title{Towards Accurate Forecasting of Renewable Energy: Building Datasets and Benchmarking Machine Learning Models for Solar and Wind Power in France}

\date{} 					% Or removing it

\author{\href{https://orcid.org/0009-0003-3947-2101}{\includegraphics[scale=0.06]{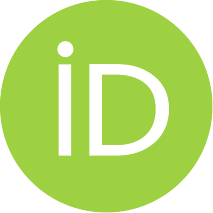}\hspace{1mm}Eloi LINDAS}\textsuperscript{1,3}\thanks{Corresponding author : \href{mailto:eloi.lindas@lsce.ipsl.fr}{eloi.lindas@lsce.ipsl.fr}}\hspace{3cm} Yannig GOUDE\textsuperscript{2,4} \hspace{3cm} Philippe CIAIS\textsuperscript{3}\\
\\
\textsuperscript{1}Atos Inno'Lab TS Bezons, Bezons, France\\
\textsuperscript{2}EDF R\&D Lab, EDF, Palaiseau, France\\
\textsuperscript{3}Laboratoire des Sciences du Climat \& de l'Environnement, IPSL, CEA/CNRS/UVSQ, \\
Université Paris-Saclay, Gif-sur-Yvette, France\\
\textsuperscript{4}Laboratoire de Mathématiques d'Orsay, CNRS, Université Paris-Saclay, Orsay, France}
\renewcommand{\headeright}{Preprint}
\renewcommand{\undertitle}{Preprint}
\renewcommand{\shorttitle}{Towards Accurate Forecasting of Renewable Energy}

\hypersetup{
pdftitle={Towards Accurate Forecasting of Renewable Energy: Building Datasets and Benchmarking Machine Learning Models for Solar and Wind Power in France},
pdfsubject={cs.LG, stat.ML},
pdfauthor={Eloi LINDAS, Yannig GOUDE, Philippe CIAIS},
pdfkeywords={electricity production, renewable sources, forecasting, machine learning},
}

\begin{document}
\maketitle
\setcounter{footnote}{0}
\begin{abstract}
Accurate prediction of non-dispatchable renewable energy sources is essential for grid stability and price prediction. Regional power supply forecasts are usually indirect through a bottom-up approach of plant-level forecasts, incorporate lagged power values, and do not use the potential of spatially resolved data. This study presents a comprehensive methodology for predicting solar and wind power production at country scale in France using machine learning models trained with spatially explicit weather data combined with spatial information about production sites' capacity. A dataset is built spanning from 2012 to 2023, using daily power production data from RTE (the national grid operator) as the target variable, with daily weather data from ERA5, production sites capacity and location, and electricity prices as input features. Three modeling approaches are explored to handle spatially resolved weather data: spatial averaging over the country, dimension reduction through principal component analysis, and a computer vision architecture to exploit complex spatial relationships. The study benchmarks state-of-the-art machine learning models as well as hyperparameter tuning approaches based on cross-validation methods on daily power production data. Results indicate that cross-validation tailored to time series is best suited to reach low error. We found that neural networks tend to outperform traditional tree-based models, which face challenges in extrapolation due to the increasing renewable capacity over time. Model performance ranges from \SI{4}{\percent} to \SI{10}{\percent} in nRMSE for midterm horizon, achieving similar error metrics to local models established at a single-plant level, highlighting the potential of these methods for regional power supply forecasting.
\end{abstract}

% keywords can be removed
\keywords{electricity production \and renewable sources \and forecasting \and machine learning}
\paragraph{Data Availability} The datasets built for this work can be accessed : \small\url{https://doi.org/10.5281/zenodo.14287949}\normalsize

\section{Introduction}
To meet the 2050 net-zero scenario \cite{paris_agreement} of the European Union (EU) reinforced by the European Green Deal which aims at decreasing net greenhouse gas emissions by \SI{55}{\percent} by 2030 \cite{european_green_deal}, sustainable energy sources have become key to clean power production and reduced emissions from the energy sector in Europe. As power demand increases, however, fossil reliance is still high, accounting for \SI{68}{\percent} of the global primary energy consumed in 2023 and \SI{40}{\percent} of the electricity produced in the European Union (EU) \cite{bp_energy_outlook, owid-electricity-mix}. Electrification coupled with more renewable and other low-carbon power supplies is needed to reduce dependence on fossil fuels. To meet the CO\textsubscript{2} emissions goals of the EU, solar and wind power generation need to double their capacity by 2030 to produce \SI{48}{\percent} of Europe's energy share \cite{irena_outlook}. \\

\noindent France has set a reduction of \SI{33}{\percent} of its emissions by 2030 compared to 1990, and pledged to reach greenhouse gas neutrality in 2050 \cite{snbc}. This involves an increase in renewable power capacity installed throughout the country. The capacity of solar and wind power plants has tripled since 2012, and this growth is expected to accelerate with the capacity being planned to double from 2017 to 2028 \cite{ppe}. Increasing renewable capacity comes with grid distribution challenges to prevent gaps between supply and demand, especially during the day when production may exceed consumption~\cite{liu_climate_2023}. Accurate forecasts of power generation can improve the stability, reliability, quality, and penetration level of renewable energy \cite{irena_innovation_brief}. Solar and wind power sources depend on environmental and climate variables such as temperature, solar radiation, and wind speed, making their load highly variable \cite{ engeland_2017, wang_exploring_2019}. This variability leads to obstacles for grid operators as they need to constantly balance the demand with the supply. This is one of the reasons why specific models for understanding and predicting day-to-day renewable power generation have motivated interest from researchers and practitioners. \\

\noindent Many studies addressed the problem of short (\SI{10}{\minute} - \SI{1}{\hour}) to medium-term (\SI{3}{\hour} - \SI{3}{ days}) forecasting of renewable power using weather data from stations or numerical weather predictions (NWP). The impact of weather data and variable importance on forecasting energy supply, PV, and wind power was studied thoroughly \cite{ de_giorgi_photovoltaic_2014, laibao_2023, vladislavleva_predicting_2011, zhong_short-term_2020}. At local scale, Malvoni \textit{et al.} used solar radiation and temperature to predict the generation of a Mediterranean PV plant \cite{malvoni_data_2016}. The effect of various climate throughout the planet on hourly PV production was also investigated by Alcaniz \textit{et al.} \cite{alcaniz_effect_2023}. Other works made use of weather forecasts such as Ahmad \textit{et al.} to maximize hydropower generated from dams \cite{ahmad_maximizing_2020}, and by Couto \textit{et al.} who examined model-based predictive features for wind power predictions \cite{couto_enhancing_2022}. Frequently, the availability of accurate weather observation is a bottleneck when working with a dedicated local area, not to mention their inherent sparsity and noise level, leading to NWP being preferred by researchers. Yet, when both types of weather data are available, they can be combined \cite{lopez_gomez_photovoltaic_2020, sharma_predicting_2011}.\\

\noindent Recent advances in forecasting variable renewable energy generation have seen statistical, machine learning, and deep learning models gain popularity among practitioners \cite{iheanetu_solar_2022, krechowicz_2022, tsai_review_2023, wang_review_2019}. Thanks to the increase in weather and power data availability and quality, models have proven to be useful in revealing driving factors and learning from complex patterns \cite{sweeney_2020}. Depending on the spatial and temporal scale, statistical models can outperform traditional physics-based models, which motivated the development of hybrid models \cite{bellinguer_2020, castillo_2023, gijon_prediction_2023}. The link function between weather conditions and PV panels or wind turbines power output has been thoroughly investigated through different types of models \cite{bilendo_2023, dolara_2015, mayer_2021, zhou_2022}. Still, challenges remain when developing models for a large region or country.\\

\noindent Statistical data-driven models such as auto-regressive moving average (ARMA) and their variant (ARIMA, SARIMAX ...) have demonstrated reasonable performance, as shown in recent work \cite{chen_wind_2018, ryu_evaluation_2022}. Support vector machine (SVM), k-nearest neighbors (kNN), generalized additive models (GAM), tree-based and boosted models also gave good performance in forecasting power output from weather data~\cite{condemi_hydro-power_2021, kim_two-step_2019}. Current trends have seen the use of artificial neural networks (ANN), computer vision (CV) and natural language processing (NLP) models. Their application in renewable power forecasting shows promising performance. Multi-layered perceptrons (MLP), convolutional neural networks (CNN), vision transformers (ViT) \cite{keisler_winddragon_2024, lim_solar_2022} and sequence architectures such as recurrent neural networks (RNN) or long-short term memory deep learning models (LSTM) were also applied in various renewable energy forecasting frameworks (solar and wind) \cite{abdul_baseer_electrical_2023, elsaraiti_solar_2022}. A key advantage is their flexibility and ability to combine several data sources to make predictions, not to mention the different ways they can exploit complex spatiotemporal data.\\ 

\noindent Research on statistical models is not limited to model architectures. Data pre-processing techniques are also important to improve forecast performance. Principal components analysis (PCA), wavelet decomposition (WD), time series detrending, and exponential smoothing can be applied to extract relevant features, reduce dimension, remove noise, or reveal pertinent phenomena from the data~\cite{iheanetu_solar_2022, liu_data_2019}. These techniques are mainly used as a first step to improve the robustness and performance of a model. It is important to point out that such techniques can be applied regardless of the type of data at hand, whether it is time series or gridded data over a region albeit the second option being less explored.\\

\noindent Besides the methodology and models used for forecasting, differences between studies arise from the input and output data. Depending on the purpose and the availability of the data, the time and space resolution as well as temporal and spatial ranges differ between studies \cite{engeland_2017}. Research works encompass scales from short-term single plant forecasts with a time resolution of 5-10 minutes \cite{gijon_prediction_2023, malvoni_forecasting_2017, ryu_evaluation_2022} to medium-term daily forecasts of a region \cite{kim_daily_2017}. However, due to the lack of available good quality data, regional forecasts are often made out of single plant forecasts aggregated to the desired region. This means an indirect prediction of regional power supply. Moreover, the temporal scale rarely exceeds a few years' worth of data \cite{chen_wind_2018, iheanetu_solar_2022}. Thus, gaps exist between short to medium-term and regional forecasts leading to difficulties in comparing results between studies and improving modeling performance. \\

\noindent Most prior studies have used a bottom-up approach based on single-plant models, which neglects the integration of spatial information for prediction. Additionally, many existing models enhanced their performance by incorporating lagged data of the target time series itself, such as power supply from the previous day or hour. To overcome these limitations, in this study, we use supervised machine learning models and test the impact of using spatially resolved data as model inputs. We also decided to exclude the use of lagged inputs from time series themselves as model inputs. The first goal is to assess the influence of model calibration procedure, especially the cross-validation protocol, on time series-based model error estimation. The second goal is to compare models ingesting explicit weather "images" against averaged variables as inputs.\\ 

\noindent We first explain how we build input datasets for wind and PV production integrating spatially resolved weather data and generation units capacity and locations. These input images span the period from 2012-01-01 to 2023-12-31 at hourly resolution as presented in section \ref{sec:data_section}. Secondly, we present three different modeling approaches to handle the weather-gridded data to forecast daily wind and PV power production in section \ref{sec:model_approach}. Finally, we explore cross-validation and hyperparameters optimization procedures in section \ref{sec:cv_hpo} to give insights and recommendations for model calibration before benchmarking widespread state-of-the-art machine learning models on our different modeling approaches in section~\ref{sec:results_section}.

\section{Data}\label{sec:data_section}
\begin{figure}[ht!]
    \centering
    \includegraphics[width=0.9\textwidth]{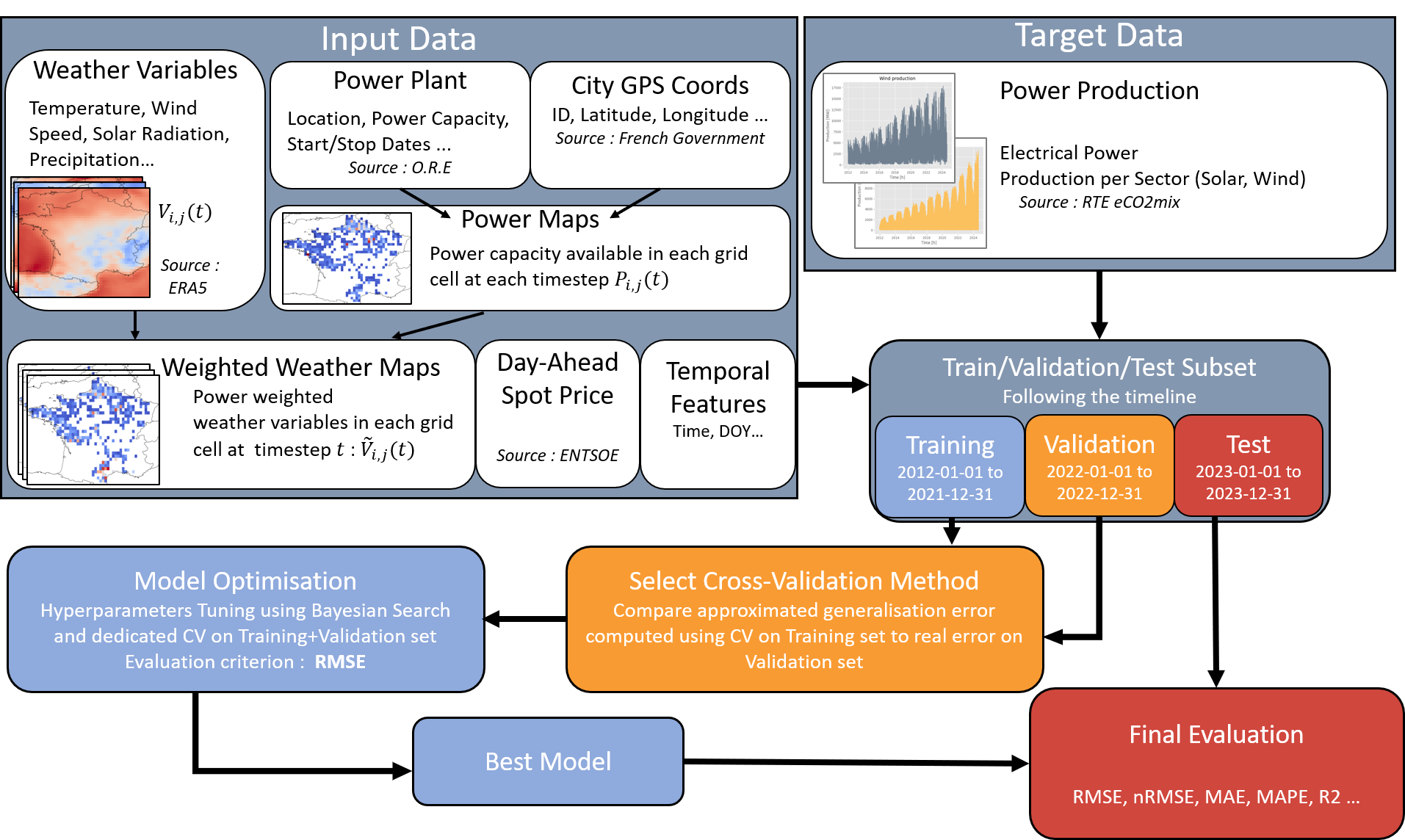}
    \caption{Global framework of this study represented schematically}
    \label{fig:data_processing_workflow}
\end{figure}

In this section, we describe the target power supply data, the input weather data and power units data, and other input data sources, with the processing workflow to prepare them as input of supervised learning approaches. Figure \ref{fig:data_processing_workflow} presents the overall approach, with more details given in the following sections. 

\subsection{Target Data}
We used as target Wind and Solar power from the RTE eCO\textsubscript{2}mix database. RTE is the public French national Transmission System Operator (TSO) managing the whole electrical grid. RTE provides near real-time data on electrical consumption, production, flows, and CO\textsubscript{2} emissions within the eCO\textsubscript{2}mix application\footnote{RTE eCO\textsubscript{2}mix website: \url{https://www.rte-france.com/en/eco2mix}. Accessed : 2024-09-19}. Electricity production data from RTE covers $8$ sectors: Coal, Oil, Gas, Nuclear, Hydro, Solar, Wind, and Bioenergy. We recovered production data for non-dispatchable renewable Wind and Solar power. Solar refers to photovoltaic solar panels and Wind to both onshore and offshore turbines.\\ 

\noindent Time-wise, data is available since 2012-01-01 and was retrieved until 2023-12-31. Resolution is half hourly from 2012-01-01 to 2023-01-31 and quarter-hourly from 2023-02-01 to 2023-12-31\footnote{Resolutions might change for 2023 in future releases. Current resolutions and types of data are given for the September 2024 release}. We aggregated the data to hourly resolution to be consistent with the time resolution of our inputs (see section \ref{sec:input_data}). Data being available at country (NUTS0) or regional (NUTS1) scale, we chose to work directly with country scale data. This dataset excluded Corsica and other french islands or overseas territories which are considered self-sufficient in electricity.\\

\begin{figure}[ht!]
    \centering
    \includegraphics[width=0.9\textwidth]{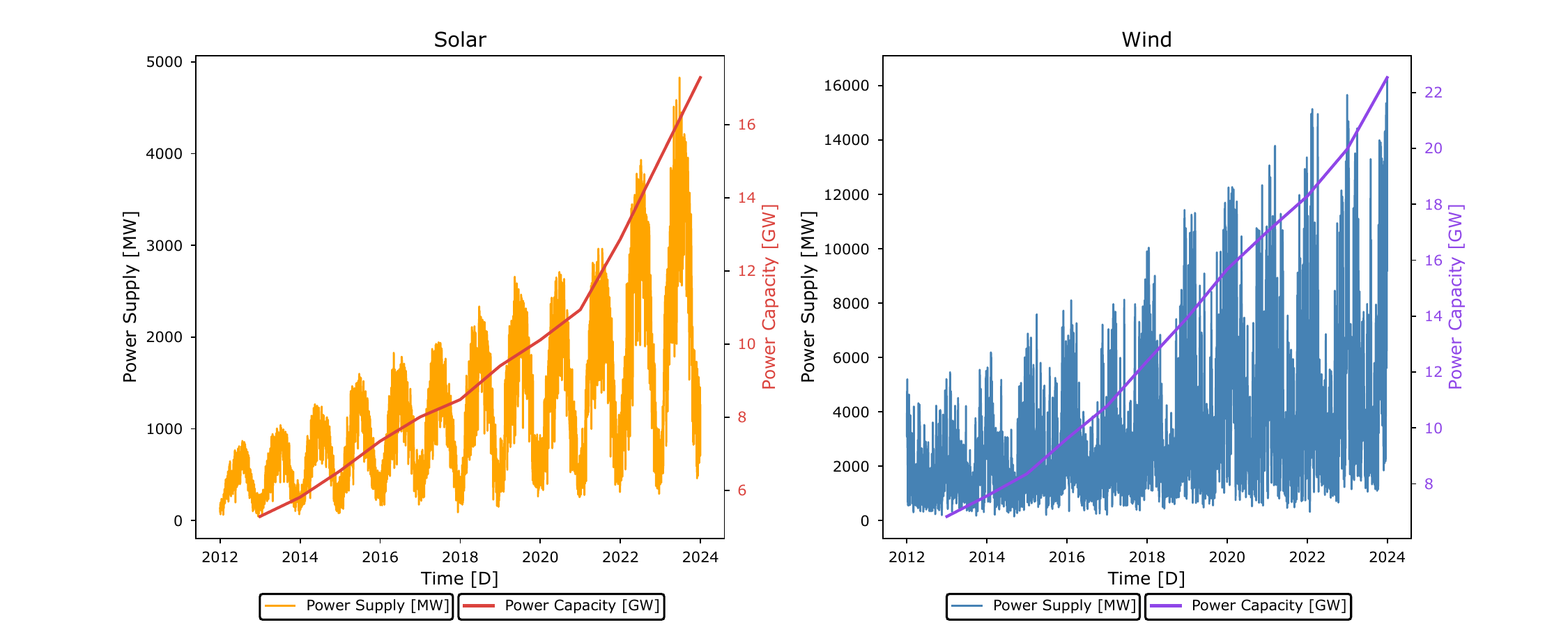}
    \caption{Power supply and capacity time series for Wind and Solar in France for the period of interest. The power capacity curves have been smoothed to yearly resolution}
    \label{fig:renewables_plot}
\end{figure}

\noindent France is part of the European Union electricity market and EU grid interconnection. In this work, we aim to model the electrical power produced using Solar and Wind from France only without taking into account any connection with neighboring countries. Therefore, we did not integrate imports and exports into our power supply target and retained only the production data, presented in figure \ref{fig:renewables_plot}.

\subsection{Input Data}\label{sec:input_data}
Our input data is based on gridded weather data weighted by the power capacity available at the given time and location, electricity day-ahead spot price, and other temporal features such as time or day of the year. We combined several different high-quality open-access databases from French governmental or government-affiliated organisms to create coherent inputs.

\subsubsection{Weather Data}
\noindent We recovered hourly weather data from the ERA5 reanalysis \cite{hersbach_era5_2020} on single levels for the period of interest from 2012-01-01 to 2023-12-31. We used the domain bounded by \SI{51}{\degree} North, \SI{42.5}{\degree} South, \SI{-4.55}{\degree} West, and \SI{7.95}{\degree} East which covers France, re-interpolating the original spatial grrid of \SI{0.25}{\degree}~$\times$~\SI{0.25}{\degree} or \SI{30}{km}~$\times$~\SI{30}{km}. The weather variables we selected are those usually used for renewable power prediction: temperature at 2 meters, Northward and Eastward wind speed at 10 meters and at 100 meters, instantaneous wind gust speed at 10 meters, surface solar radiation downwards, total precipitation, evaporation, and runoff (table~\ref{tab:climate_variables}). To select the variables relevant to Wind and Solar power, we used the mutual information between weather variables and power supply targets \cite{kraskov_2004}. We normalized the mutual information to 1 and kept only variables that had a score higher than \SI{20}{\percent}. This leads to hourly maps with 35 latitude and 51 longitude points for each considered variable in netCDF files.

\subsubsection{Power Units Location, Capacity \& Activity}
To get information on the location of facilities with installed solar panels or wind turbines, we used yearly released data from the Opérateurs Réseaux Energies (O.R.E)\footnote{Dataset used can be retrieved from O.R.E website: \href{https://opendata.agenceore.fr/explore/dataset/registre-national-installation-production-stockage-electricite-agrege-311223/information/?stage_theme=true&disjunctive.epci&disjunctive.departement&disjunctive.region&disjunctive.filiere&disjunctive.combustible&disjunctive.combustiblessecondaires&disjunctive.technologie&disjunctive.regime&disjunctive.gestionnaire}{https://opendata.agenceore.fr/pages/accueil/}} agency database of all electrical facilities used for producing or storing electricity in France. The inventory published on 2023-12-31 contained around 84,000 electricity-producing units amongst which 2,183 are wind facilities and 72,703 are PV farms. Rooftop PV panels dedicated to auto-consumption are not included. Because the ORE dataset did not provide the exact location of each facility, we merged it with the French governmental city database\footnote{This database can be found on the French government Open-Data platform: \href{https://data.enseignementsup-recherche.gouv.fr/explore/dataset/fr-esr-referentiel-geographique}{https://data.enseignementsup-recherche.gouv.fr/explore/dataset/fr-esr-referentiel-geographique/export/}} using City ID, to allocate each facility to a 30 km grid cell of our weather maps . A city refers to a NUTS 4 entity. City ID is a unique identifier provided to every French city by Institut National de la Statistique et des Etudes Economiques (INSEE). Facilities' city IDs that were missing in O.R.E accounted for less than \SI{2}{\percent} of the data and were discarded. We assigned facilities to their corresponding Wind or Solar sector, keeping only PV panels for Solar and including both offshore and onshore turbines for Wind. The maximum power that can be produced by each facility in \SI{}{\mega\watt} provided by O.R.E was used as its capacity. Some power capacity data were missing, representing \SI{0.25}{\percent} fo the data and thus were discarded. To account for the activity period of each facility, we added its start and stop dates. If the stop date was not given in the O.R.E inventory we assumed that the facility was still in activity. For the start date, we used the start-up date or the date the plant was connected to the grid. We verified that those two starting dates were close to each other for facilities where both were reported. After latitude, longitude, sector, power capacity, and start/stop dates for each facility were added, we only dropped \SI{4.4}{\percent} of the initial O.R.E dataset. Most of those discarded plants are located overseas or in Corsica.

\subsubsection{Power Weighted Weather Maps}
\begin{figure}[ht!]
    \centering  
    \includegraphics[width=0.8\textwidth]{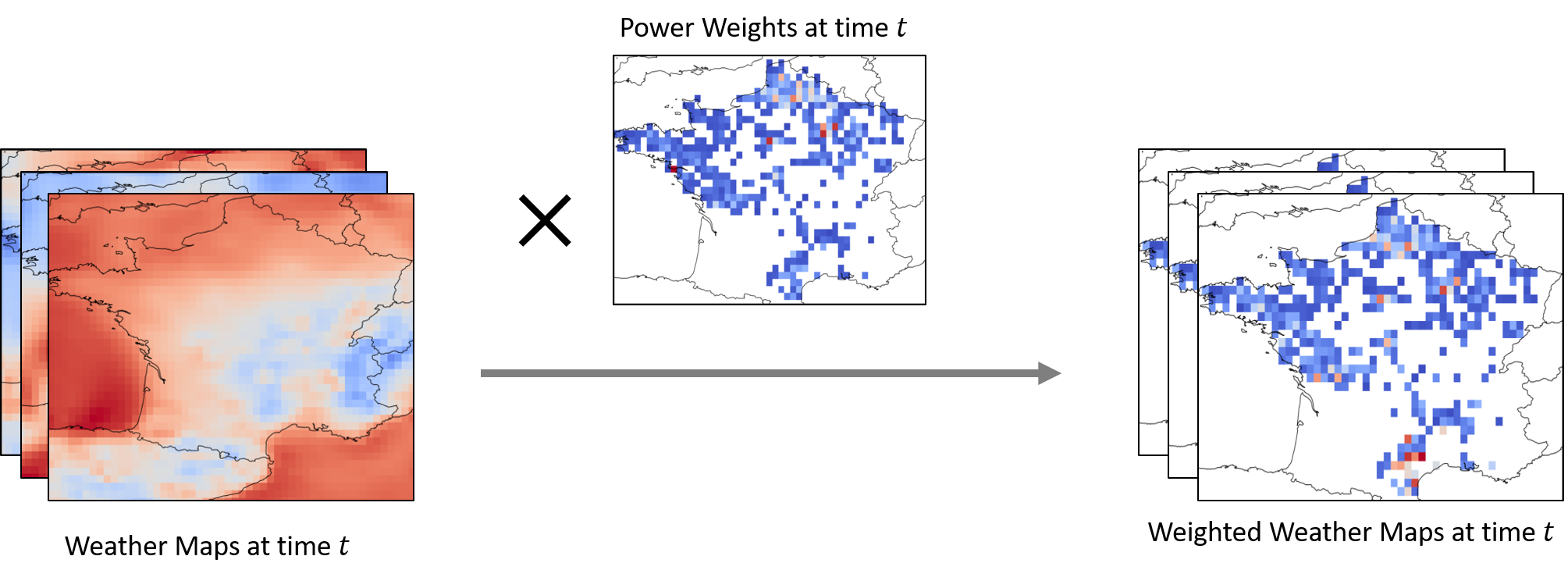}
    \caption{Illustration of power-weighted weather maps creation for Wind}
    \label{fig:final_input_creation}
\end{figure}

We generated power capacity-weighted weather maps, by assigning each power facility to the nearest grid cell in the gridded hourly weather data. The weather parameters are thus multiplied by the power capacity weights defined as :
\begin{equation}
  w_{i,j}^{t} = \dfrac{P_{i,j}^{t}}{\sum_{t}\sum_{i,j}P_{i,j}^{t}}
\end{equation}
with the power capacity $P_{i,j}^{t}$ at time $t$ and latitude, longitude $i,j$ in \SI{}{\mega\watt}. We use a spatiotemporal normalization of the weights to account for the fact that non-dispatchable renewable energy sources have seen their available production capacity increase in the last few years (see Figure \ref{fig:renewables_plot}). Since this behavior is expected to carry on, it is important to account for it in the model's input. Figure \ref{fig:final_input_creation} recaps the weighted weather map creation schematically.\\

\subsubsection{Additional Input Features}
To ensure that models could grasp all of the seasonality and trend, we added two temporal features as it is usually done in the electricity forecasting literature \cite{chatfield_1986, goude_2016, taylor_2010}. The time step converted to a numerical integer, and the day of the year encoded using a cosine: $doy_{cos} = \cos \left(\dfrac{2\pi doy_{int}}{365} \right)$ where $doy_{int}$ is the day of the year encoded as an integer between 1 and 365. We used those two temporal features for the Wind and Solar sectors. However, to be more consistent with the physical process of producing electricity with PV panels we replaced $doy_{cos}$ for Solar by the sunshine duration of the day. This duration was computed from sunrise and sunset times. We did it for every grid cell and timestep. \\

\noindent Even though PV and Wind power supply to the grid are related to weather conditions, they are also dependent on the demand that electricity providers need to meet. The last few years have seen negative electricity prices on the market soar as the electrical demand was low and the available renewable power was in oversupply. This led to a new practice from electricity providers called curtailment which consists of deliberately restricting the electricity generation from renewable energy sources to prevent negative prices \cite{bieber_2022, devita_2018, yasuda_2022}. Thus, we added as input the electricity spot price for France at hourly resolution from ENTSO-E\footnote{ENTSO-E Transparency Platform: \url{https://transparency.entsoe.eu/}}. There are different ways participants trade electricity on the market and therefore different electricity prices. We chose to use the auction day-ahead spot price as it is the only one that can be freely retrieved through ENTSO-E. Auction day-ahead spot price is the price of a \SI{}{\mega\watt\per\hour} decided the day before delivery through an auction.\\ 

\noindent The above-described data processing methodology and workflow allowed us to have input and target datasets for Solar and Wind power, designed for a supervised learning approach, and consisting of a set of $(X, Y)$ observations. $X$ refers to hourly weather maps gridded over France for each selected weather variable weighted by the power capacity of plants located in the corresponding cells. It also includes day-ahead spot price and temporal features such as the time and day of year or sunshine duration. $Y$ refers to the corresponding electrical power produced during this hour. There are 110808 hourly observations for $(X, Y)$ spanning 4383 days with a 35 $\times$ 51 grid for each time step.

\section{Models \& Calibration}
This section describes the models we tested to predict electricity power production from weather variables. It also includes a discussion on model calibration techniques. 

\subsection{Modeling Choices \& Approaches}\label{sec:model_approach}
As our aim is to develop models able to predict the power production of PV and Wind for a day given the weather conditions, day-ahead price, and temporal features of that same day, we aggregated all input data from hourly to daily resolution. Aggregation also helped to increase the signal-to-noise ratio and prevent over-fitting when predicting daily power from hourly data. This leads to a day-to-day prediction approach without utilizing values of the previous days. On operation, real forecasts could then be easily obtained with our model by plugging daily weather forecasts from numerical weather prediction models. \\

\subsubsection{Models Architectures}
\begin{figure}[ht!]
    \centering
    \includegraphics[width=0.8\textwidth]{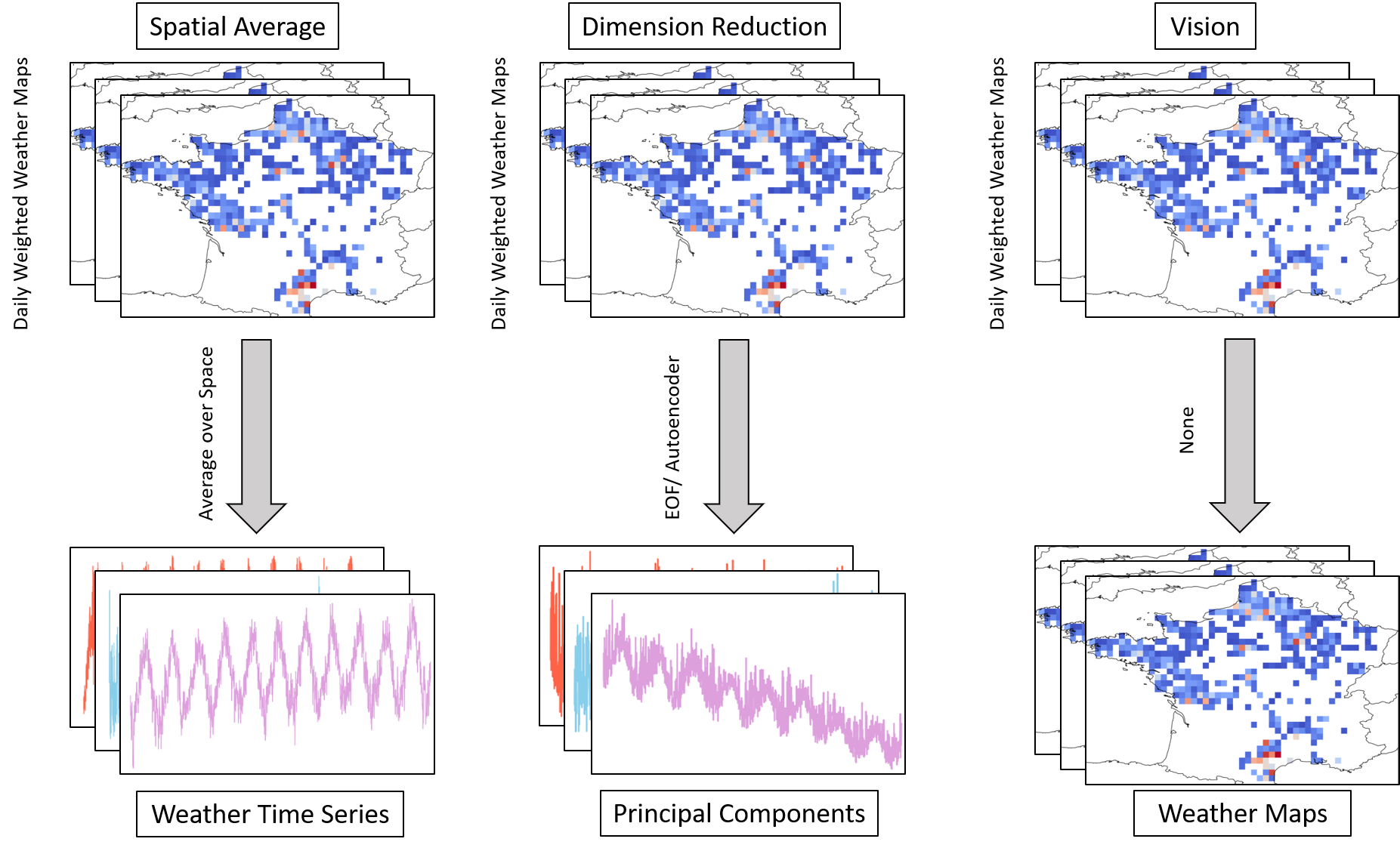}
    \caption{Representation of the 3 modeling approaches used in this work to make use of weather maps}
    \label{fig:modeling_approaches}
\end{figure}

We chose to test three modeling architectures of increasing complexity, summarized in figure \ref{fig:modeling_approaches}: first using power-weighted weather images averaged over the whole French territory, second applying to power-weighted weather a dimension reduction method, and third applying a vision or image-based technique. 

\paragraph{Models using Spatially Averaged Images as Input}
\hfill \break The first approach is to train models on spatially-averaged input data, to have a time series-to-time series regression framework. After averaging, weather time series are combined with price and temporal features series to leverage one-to-one models (models using one input point to predict the corresponding target point). In this family of models, we tested linear regressions, generalized additive models (GAM), tree-based models, boosting or artificial neural networks, all proven to be capable of reaching state-of-the-art performance \cite{ gang_2023, gaillard_2016, krechowicz_2022, laibao_2023, wood_2014}.

\paragraph{Models using Dimensionally Reduced Input Images}
\hfill \break The second approach is to use dimension reduction techniques to extract key features from our high-dimensional input power-weighted weather maps before combining them with price and other time features for training a model \cite{teste_2024}. Several dimension reduction methods exist, ranging from Empirical Orthogonal Function (EOF), widely used in the earth sciences community, to auto-encoder (AE) based on deep network architectures. These methods enable us to reduce the dimension of the input space, yet providing rich features. In this work we focused on PCA and optimized the number of principal components as any other model hyperparameter. After obtaining the principal components which behave as time-series we applied the same models as for the spatial average: tree-based models, GAM, neural networks~...

\paragraph{Models using Images as Input}
\hfill \break The third approach consists of building models capable of directly ingesting the power-weighted weather maps alongside with price, and temporal features. Here, we used a CNN architecture, previously shown to be capable in image classification, segmentation or regression tasks even though they are now slowly being replaced by better performing ViTs \cite{keisler_winddragon_2024}. \\

\subsection{Train, Validation \& Test subsets}
We split our dataset into a training and a test subset for the evaluation of model performance. As our data is time-dependent, power production changed throughout the years mainly due to openings of new facilities, we chose the period from 2012-01-01 to 2022-12-31 to be the train set and 2023-01-01 to 2023-12-31 to be the test set. Nonetheless, hyperparameter tuning is a key step of model development as it often makes the difference between poor and high-performing models. To perform hyperparameter optimization (HPO) we can use different cross-validation methods as well as different optimization frameworks. To ensure the robustness of our model selection procedure, we chose to keep a validation set dedicated to the investigation of cross-validation and optimization methods. This validation set spans the period from 2022-01-01 to 2022-12-31. After choosing a proper model selection and HPO procedure it is included in the train set for final HPO and model calibration before evaluation on the test set, as described below.

\subsection{Cross-Validation \& Hyperparameters Optimization}\label{sec:cv_hpo}

Cross-validation is used to approximate the generalization error i.e. the error of the trained model exposed to new unseen data \cite{hyndman}. Different techniques are used for splitting the training set into a new training set to train the model and a new left-out test set to evaluate its performance for computing the approximated generalization error. This step is usually combined with HPO to select the best set of hyperparameters for a given model architecture. Selecting the best-suited calibration procedure is a complicated process \cite{arlot_2009, bergstra_algorithms_nodate} and we explain below the proposed optimization scheme.\\

\subsubsection{Procedures Inspected}

\noindent Our data is time-dependent since our target is a power supply time series. Different studies investigated which cross-validation procedure was best suited in this case \cite{bergmeir_2012, cerqueira_2019, tashman_2000}. However, the scope of those studies was mainly synthetic and stationary not to mention small, i.e. a few hundred points, time series. Another major limitation is that even though real datasets were used, those modeling approaches involved lagged values of the target time series as predictors, which is excluded in our case. Therefore, we chose to study different cross-validation procedures and HPO algorithms to guide the choices for the calibration of our models. We did these experiments using only the models based on spatial averages of input weather images. The following cross-validation procedures were used : 
\begin{itemize}
  \item \textbf{Hold-Out}: Split the training set into a train set and a test set.
  \item \textbf{K-Fold}: Split the training set into $K$ folds. At each iteration, a fold is chosen to be the test set while the $K-1$ others form the train set. Iterate until all folds were used as test once. After all the iterations, the approximated generalization error is taken to be the average of the error made on each test fold.
  \item \textbf{Expanding}: Split the training set into $K$ folds following the order of the samples. During the $i^{th}$ iteration, the first $i$ folds are used as the train set and the $i+1$ fold is used as the test. Repeat until the entire training set has been used. After all the iterations, the approximated generalization error is taken to be the average of the error made on each test folds
  \item \textbf{Sliding}: Split the training set into $K$ folds following the order of the samples. During the $i^{th}$ iteration, the $i$ fold is used as the train set, and the $i+1$ fold is used as the test. Repeat until the entire training set has been used. After all the iterations, the approximated generalization error is taken to be the average of the error made on each test folds
  \item \textbf{Blocking}: Choose a block length $l$ based on the temporal structure to conserve most of the correlation between neighboring samples. Split training set into blocks of length $l$. Attribute blocks to the train or test set at random. Inspired from \cite{wood_2024}.
\end{itemize}

\begin{figure}[ht!]
    \centering
    \includegraphics[width=0.8\textwidth]{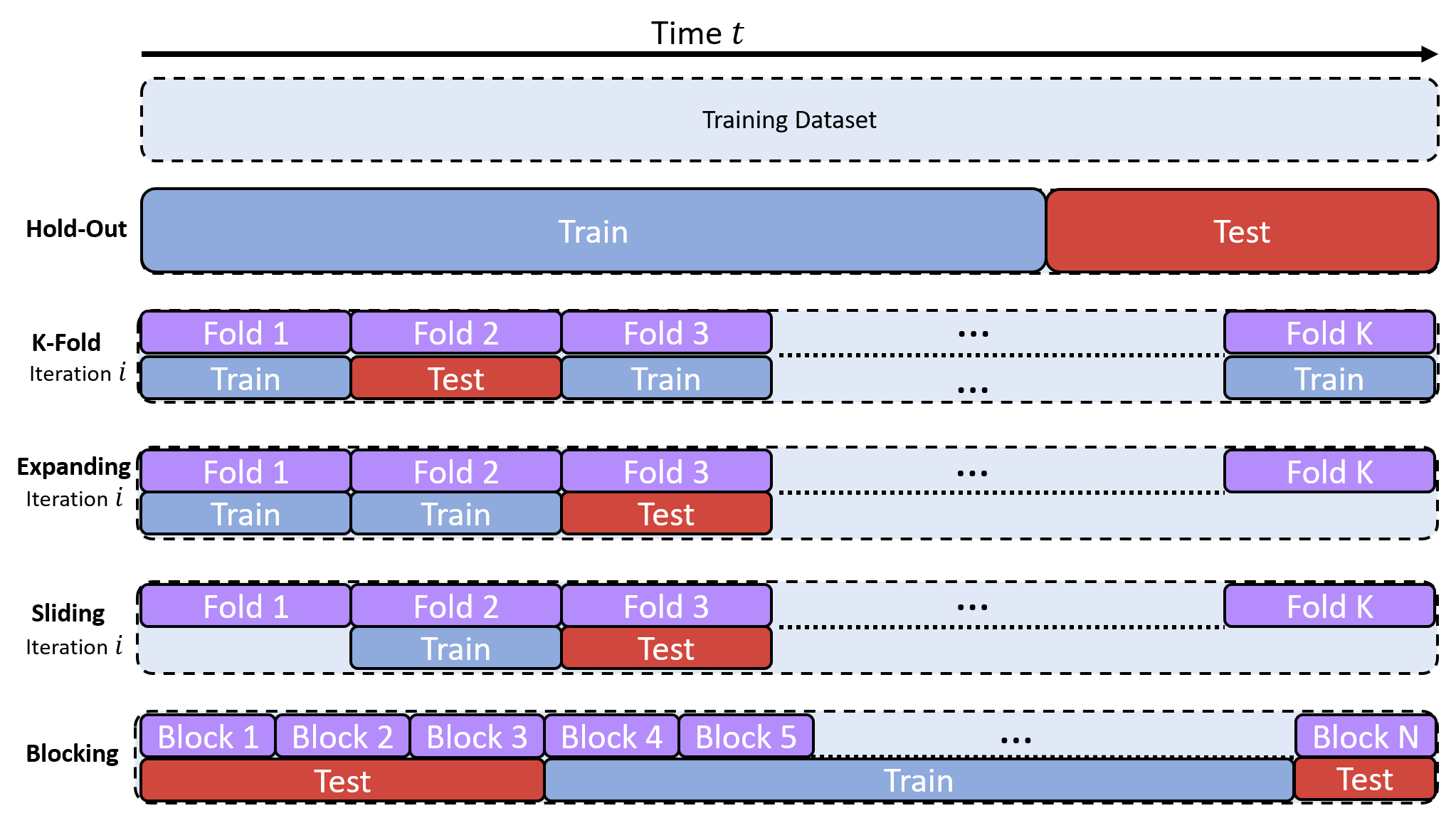}
    \caption{Different cross-validation procedures considered in this work represented schematically. For Hold-Out and K-Fold only the method without prior random shuffling are represented}
    \label{fig:cv_scheme}
\end{figure}

\noindent Figure \ref{fig:cv_scheme} shows the scheme of these five cross-validation methods. We split the data into a one-year test set for the Hold-Out method, \SI{10}{} splits to get yearly folds for every method using folds and blocks of \SI{7}{} days for the blocking method. The block size was chosen to keep most of the temporal structure using autocorrelation and partial autocorrelation analysis. We also considered the shuffling variants of the K-Fold and Hold-Out methods which involve randomly shuffling the samples before the folds or subset attributions.\\

\noindent Regarding hyperparameter optimization, we compared two optimization algorithms: Random search and Bayesian search using Gaussian Processes \cite{bischl_hyperparameter_2023, bergstra_algorithms_nodate}.\\
To assess the impacts of cross-validation and HPO for different model architectures, we repeated the experiments using three models: a random forest, a tree-based boosting scheme (XGBoost), and a feed-forward neural network or MLP. In total, this led to 7 cross-validations x 2 HPO x 3 models estimators of the generalization error. At first glance, one might think that cross-validation procedures that respect the temporal order of the data are best suited to our approach. Still, we wanted to make an informed decision by doing the experiments. Our final goal is to choose the pairs of cross-validation techniques and HPO algorithms that give the 'best' estimator of the generalization error. Here 'best' refers to different criteria ranging from the precision of the generalization error estimate to the computational resource usage. \\ 

\subsubsection{Cross-Validation Experiments}
\noindent As cross-validation's main goal is to obtain an approximate of the generalization error $\hat{\varepsilon}$ we monitored how far the estimate was from the real error. To do so, we recorded for each of the 100 optimization iterations the test error made during cross-validation on the training part of the data for a given set of hyperparameters. Then we compared it to the real generalization error $\varepsilon$ made on the validation set. Here the training and validation part refers to the one visible in figure \ref{fig:data_processing_workflow}. Since we are dealing with a regression task, the error $\varepsilon$ was taken to be the root mean squared error (RMSE) of the modeled and observed daily power production. See appendix \ref{appendixB} for metrics definition. Our target being a power production daily time series, the unit of RMSE is \SI{}{\mega \watt}. Given the real generalization error $\varepsilon$ and its estimate $\hat{\varepsilon}$ from cross-validation, for each procedure we computed the difference between the two quantities as $\Delta \varepsilon = \varepsilon - \hat{\varepsilon}$ and analyzed the average $\overline{\Delta \varepsilon}$ and its standard deviation $\sigma(\Delta \varepsilon)$ across the HPs. We also determined the optimum value of $\hat{\varepsilon}$ reach after optimization and compared it with the real error in $\Delta \varepsilon_{min}$.\\

\noindent During the experiments we monitored the time taken to perform one iteration and the permutation feature importance of each feature obtained during cross-validation compared to the one obtained on the validation set. These times of computation tell us how costly each error estimation method was. The feature importance tells us if the cross-validation technique impacted the interpretability of the model. Last, we experimented with different dataset sizes to inspect the influence of data size on cross-validation methods since the literature only deals with small sample sizes. As the dataset size increases, older and older data are utilized for training. Computation times can be found in table \ref{tab:cv_times} and results for random forest on Solar are presented in figure \ref{fig:cv_solar_rf} and \ref{fig:cv_solar_rf_size}. Results for other models on Solar are in appendix \ref{appendixC} and for Wind are in appendix \ref{appendixD}. Results about permutation feature importance showed that despite the different cross-validation methods, the ranking of the features stayed the same for the different hyperparameter combinations explored, meaning that the method does not impact the model interpretability. \\

\begin{figure}[ht!]
    \centering
    \includegraphics[width=0.8\textwidth]{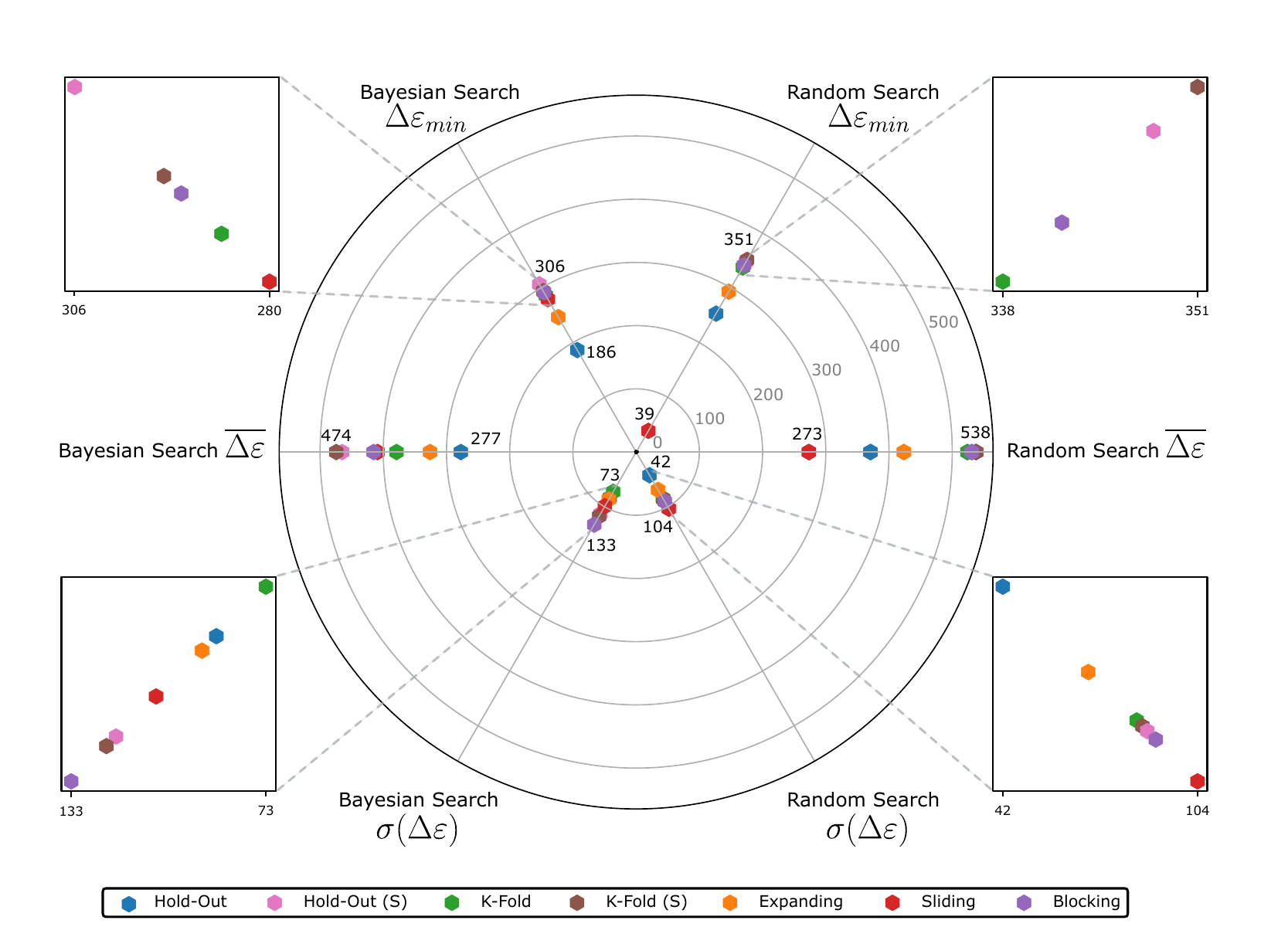}
    \caption{Results of different cross-validation techniques for random forest on Solar. Each axis represents a monitored quantity for a given HPO optimization procedure. The values for each method are plotted as points and only the worst and best values for each axis are printed. The (S) indicates the shuffling variant of the method}
    \label{fig:cv_solar_rf}
\end{figure}

\begin{figure}[ht!]
\centering
    \includegraphics[width=0.8\textwidth]{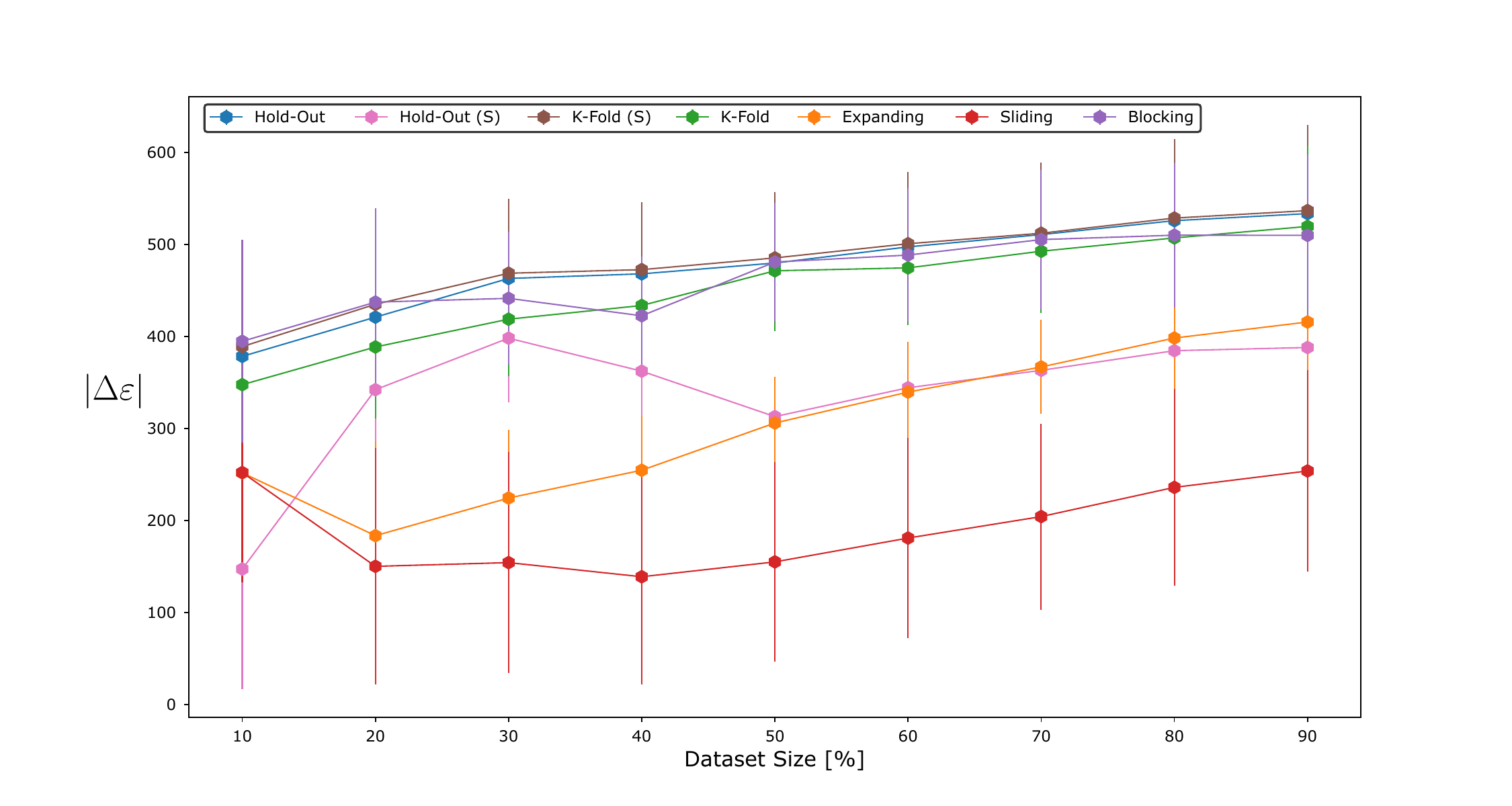}
    \caption{Robustness of cross-validation procedure regarding dataset size for random forest on Solar. The marker indicates the average $|\Delta \varepsilon|$ while the error bars display the standard deviation. The (S) indicates the shuffling variant of the method}
    \label{fig:cv_solar_rf_size}
\end{figure}

\noindent On the radar chart of figure \ref{fig:cv_solar_rf} we can see that $\Delta \varepsilon$ is positive on average and for the optimum. This means that our generalization error estimates $\hat{\varepsilon}$ is lower than the real error $\varepsilon$. In other words, the cross-validation tends to overestimate the model performance leading to overconfidence in the model. We can also see that methods that do not preserve the chronological order or shuffling perform worse than those that do. Especially Hold-Out, Expanding, and Sliding lead to the closest estimate on average and the optimum for both searches. However, Sliding is the most sensitive to the set of hyperparameters as its variability $\sigma(\Delta \varepsilon)$ is the highest. This might stem from its small training set size which never exceeds 1 year of data. This is also confirmed by the error bars of figure \ref{fig:cv_solar_rf_size}. This same figure shows that increasing the dataset size by appending older and older data leads to a slight increase in $|\Delta \varepsilon|$ meaning that our generalization error estimate is moving away from the real one. This is because older data such as 2012 carry less meaningful information than more recent data such as 2020 for predicting the validation set which is the year 2022. This behavior also explains why some methods display an inflection point for a certain dataset size meaning that there is an optimum past period of time to consider to make better predictions on the validation set. \\

\begin{table}[ht]
\resizebox{\textwidth}{!}{
\begin{tabular}{l|l|l|l|l|l|l|l|l}
\toprule
\textbf{Model} & \textbf{Sector} & \textbf{Hold-Out} & \textbf{Hold-Out (S)} & \textbf{K-Fold} & \textbf{K-Fold (S)} & \textbf{Expanding} & \textbf{Sliding} & \textbf{Blocking}\\
\midrule
Forest & Solar & 2.3 $\pm$ 1.6 \silvermedal& 2.1 $\pm$ 1.8 \goldmedal& 22.7 $\pm$ 11.8 & 14.1 $\pm$ 14 & 15.4 $\pm$ 9.3 & 19.8 $\pm$ 1.5 & 2.3 $\pm$ 1.7 \bronzemedal\\
Boosting & Solar & 3.0 $\pm$ 3.9 \silvermedal & 3.9 $\pm$ 4.5 \bronzemedal & 21.2 $\pm$ 17.8 & 53.2 $\pm$ 41.6 &33.4 $\pm$ 27.7 & 1.7 $\pm$ 1.4 \goldmedal& 4.6 $\pm$ 5.7\\
Neural Network & Solar & 2.7 $\pm$ 2.3 \goldmedal & 2.8 $\pm$ 2.4 \bronzemedal& 27.6 $\pm$ 24.1 & 27.8 $\pm$ 23.9 & 16.3 $\pm$ 14.0 & 4.3 $\pm$ 3.3& 2.7 $\pm$ 2.3 \silvermedal\\
Forest & Wind & 2.4 $\pm$ 2.4 \goldmedal & 3.8 $\pm$ 2.5 & 19.6 $\pm$ 19.1 & 37.0 $\pm$ 23.2 & 9.9 $\pm$ 10.9 & 1.9 $\pm$ 1.0 \goldmedal & 1.8 $\pm$ 1.8 \silvermedal\\
Boosting & Wind & 3.7 $\pm$ 2.6 \goldmedal& 4.5 $\pm$ 3.9 \silvermedal& 61.2 $\pm$ 78.7 & 122.6 $\pm$ 92.2 & 78.4 $\pm$ 75.5 & 6.9 $\pm$ 3.9 & 6.2 $\pm$ 3.7 \bronzemedal\\
Neural Network & Wind & 2.8 $\pm$ 2.4 \goldmedal & 2.8 $\pm$ 2.4 \silvermedal & 28.1 $\pm$ 24.4 & 57.5 $\pm$ 53.3 & 33.0 $\pm$ 30.0 & 8.7 $\pm$ 7.4 & 2.8 $\pm$ 2.4 \bronzemedal\\
\bottomrule
\end{tabular}}
\caption{Average and standard deviation of computing times for 1 iteration for each cross-validation method in seconds. The (S) indicates the shuffling variant of the method. Medals indicate the top three fastest methods for each model and dataset}
\label{tab:cv_times}
\end{table}

\noindent The same conclusions hold for boosting and feed-forward neural networks on the Solar dataset (see figures \ref{fig:cv_solar_gb}, \ref{fig:cv_solar_gb_size}, \ref{fig:cv_solar_nn} and \ref{fig:cv_solar_nn_size}). It is worth mentioning that the neural network shows a high variability and a high $\Delta \varepsilon$ for the Bayesian search HPO, suggesting that this algorithm might not be the best for optimizing neural network hyperparameters. For the Wind dataset (see appendix \ref{appendixD}), Hold-Out, Sliding, and Expanding methods are the best methods to estimate the generalization error for all 3 model architectures. Yet, we can see for the random forest and boosting models that increasing the dataset size with older data does help better approximate the generalization error with the Expanding and Sliding methods. This means that in the Wind dataset, older data still carries meaningful information for predicting the most recent validation set, even if there is a pronounced annual trend in the wind power production time series (see figure \ref{fig:renewables_plot}). \\

\noindent Lastly, table \ref{tab:cv_times} shows that cross-validation procedures involving folds are more computationally intensive per iteration, as one can expect. Combined with the previous graphs we can conclude that the longer computing times arising from the use of K-Fold methods are not worth it since Hold-Out and Sliding are better performers and between 5 to 10 times faster to compute per iteration. \\ 

\noindent From the result of those experiments testing different cross-validations, with different HPO and different model architectures we were able to make recommendations on how to choose a model selection procedure when dealing with time series to time series forecasting from covariates. We found that dedicated procedures that keep the chronological order during cross-validation perform better than standard K-Fold or shuffled Hold-Out. Depending on the model architecture and the underlying data, some techniques tend to overestimate or underestimate model performance leading to underconfidence or overconfidence in our model. This systematic work could be extended to deep learning models that directly ingest images as inputs, to also get recommendations to push their performance even further.\\

\section{Benchmark Results \& Discussions}\label{sec:results_section}
In this section, we present the results of our calibrated models on the training + validation set and evaluated on the test set. The best hyperparameters for each model were selected from the best generalization error, based on experiments from the previous section i.e. using Bayesian search with either an Expanding or Hold-Out cross-validation method depending on the model complexity to save computing time. Expanding was preferred over Sliding cross-validation due to the high sensitivity of Sliding to hyperparameter sets. We assessed the performance of the model using the RMSE, Mean Absolute Error~(MAE), Mean Absolute Percentage Error~(MAPE), Normalized Root Mean Squared Error~(nRMSE), and R2 score~(R2). The definitions of these metrics are given in appendix \ref{appendixB}. Table \ref{tab:solar_benchmark_results} contains all our results on the Solar dataset while results for Wind can be found in appendix \ref{appendixE}. \\

\begin{table}[ht]
\resizebox{\textwidth}{!}{
\begin{tabular}{l|l|c|ll|ll|ll|ll|ll}
\multicolumn{3}{c}{Metrics} & \multicolumn{2}{c}{\textbf{MAE}} & \multicolumn{2}{c}{\textbf{MAPE (\%)}} & \multicolumn{2}{c}{\textbf{RMSE}} & \multicolumn{2}{c}{\textbf{nRMSE (\%)}} & \multicolumn{2}{c}{\textbf{R2}} \\
\toprule
\textbf{Approach} & \textbf{Model} & \textbf{Detrend} & Train & Test & Train & Test & Train & Test & Train & Test & Train & Test \\
\midrule
\midrule
Average & Linear Regression & & 106 & 350 & 14.0 & 15.8 & 140 & 423 & 3.59 & 9.57 & 0.96 & 0.86 \\
Average & Random Forest & & 50.7 & 300 & 6.12 & 13.9 & 69.3 & 375 & 1.78 & 8.5 & 0.99 & 0.89 \\
Average & Random Forest & \checkmark & 57.4 & 179 \silvermedal& 6.95 & 9.83 \silvermedal & 82.3 & 279 \bronzemedal & 2.12 & 6.33 \bronzemedal & 0.99 & 0.94 \bronzemedal\\
Average & Linear Forest & & 78.0 & 300 & 9.54 & 13.9 & 109 & 374 & 2.80 & 8.45 & 0.98 & 0.89 \\
Average & Tree Boosting & & 47.3 & 253 & 6.26 & 13.9 & 63.2 & 331 & 1.63 & 7.48 & 0.99 & 0.91 \\
Average & Tree Boosting & \checkmark & 56.9 & 176 \goldmedal & 7.10 & 9.71 \goldmedal & 80.7 & 271 \goldmedal & 2.07 & 6.14 \goldmedal & 0.99 & 0.94 \goldmedal\\
Average & Linear Tree Boosting & & 106 & 352 & 14.0 & 15.8 & 140 & 425 & 3.59 & 9.62 & 0.96 & 0.86 \\
Average & GAM & & 82.0 & 321 & 10.3 & 16.0 & 113 & 401 & 2.91 & 9.10 & 0.98 & 0.87 \\
Average & MLP & & 123 & 229 & 16.4 & 11.8 & 164 & 310 & 4.21 & 7.01 & 0.95 & 0.93 \\
PCA & Linear Regression & & 141 & 350 & 17.3 & 18.2 & 190 & 435 & 4.88 & 9.85 & 0.93 & 0.85 \\
PCA & Random Forest & & 62.1 & 349 & 7.49 & 16.9 & 86.9 & 436 & 2.23 & 9.86 & 0.99 & 0.85 \\
PCA & Linear Forest & & 79.1 & 282 & 9.84 & 13.3 & 109 & 364 & 2.79 & 8.23 & 0.98 & 0.90 \\
PCA & Tree Boosting & & 53.7 & 268 & 7.43 & 14.4 & 70.3 & 381 & 1.81 & 8.63 & 0.99 & 0.89 \\
PCA & Linear Tree Boosting & & 96.9 & 319 & 12.3 & 14.5 & 133 & 403 & 3.40 & 9.12 & 0.97 & 0.87 \\
PCA & GAM & & 83.3 & 434 & 10.9 & 20.4 & 112 & 501 & 2.87 & 11.3 & 0.98 & 0.80 \\
PCA & MLP & & 85.6 & 195 & 9.52 & 10.7 & 129 & 294 & 3.33 & 6.65 & 0.97 & 0.93 \\
Vision & CNN & & 147 & 182 \bronzemedal & 15.2 & 10.1 \bronzemedal & 200 & 277 \silvermedal & 5.1 & 6.30 \silvermedal & 0.93 & 0.94 \silvermedal \\
\bottomrule
\end{tabular}}
\caption{Benchmark results for different models using 3 different modeling approaches on the Solar dataset. Medals indicate the top three best-performing models on the test set for each metric}
\label{tab:solar_benchmark_results}
\end{table}

\noindent As non-dispatchable renewables capacity increased throughout our study period, Solar and Wind power production time series have an increasing trend from 2012 to 2023 as highlighted by figure~\ref{fig:renewables_plot}. This trend requires the models to be able to extrapolate on the test set. Despite reaching state-of-the-art performance in many tasks, tree-based models such as random forest and boosting are known to face difficulties when it comes to extrapolation outside of the training domain \cite{hengl_2018, malistov_2019}. Our case makes no exception, despite low errors on the train set, random forest and boosting models errors soared on the test set (see tables~\ref{tab:solar_benchmark_results} and~\ref{tab:wind_benchmark_results}). To address this issue, many research works propose alternatives such as stochastic or linear trees \cite{gama_1999, ilic_2021, numata_2020, raymaekers_2024, haozhe_2019}. We chose to apply 2 different methods to try to solve this extrapolation problem: linear trees and detrending of the time series.\\

\noindent Our detrending scheme consisted of applying a trend estimation method such as seasonal-trend estimation using loess (STL) on the entire dataset. Once the trend is estimated, we remove it from the data. The transformed data were thus passed to the model for calibration. The predictions were obtained by reconstruction from the model's output and trend estimate. The detrending was done on both weather input and power output data as the weighting scheme introduced trends in the covariates.\\ 

\noindent Linear trees did not seem to be a silver bullet on the Solar dataset as their performance were only marginally better for the forest and worse in the case of boosting. In contrast, for the Wind dataset, they prove to be useful in enhancing the extrapolation performance. However, their performance were still far from the tree-based models predicting detrended power supply from detrended weather averages before reconstructing the proper production time series. Despite the error induced by the trend estimation and reconstruction step, this method displays some of the best results on both Solar and Wind within the spatial average method and even outside. Such behavior could be expected because the trend is estimated on the whole dataset. The extrapolation problem is weaker for GAM and MLP as they manage to better grasp the trend achieving better performance on the test set.\\ 

\noindent Compared with the spatial input averaging approach, using tree-based models with PCA did not achieve better performance due to the extracted principal components exhibiting the same trend as the spatial averages. This time we only applied linear trees as detrending principal components was more challenging. They exhibited a small improvement on the Solar dataset but a bigger decrease in performance when used to predict Wind power supply. Combining PCA with GAMs does not seem to improve performance on both datasets. For MLP it depends on the sector but one thing that we noticed after our calibration is that networks combined with PCA are deeper than networks without it, meaning that it requires more layers to extract meaningful information from the principal components.\\

\begin{figure}[ht!]
    \centering
    \includegraphics[width=0.8\textwidth]{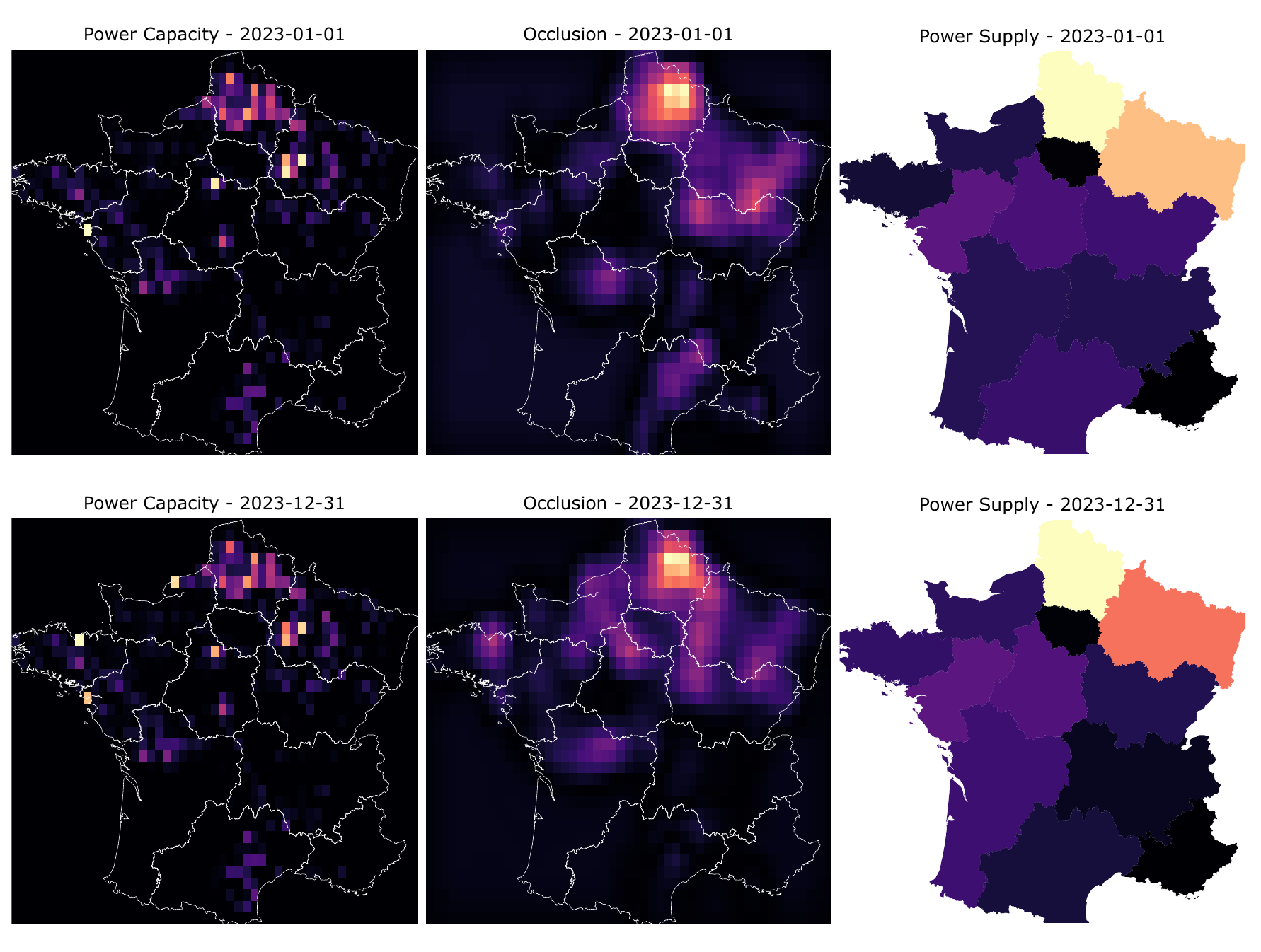}
    \caption{Power capacity, Occlusion attribution, and regional realized power supply for early and late 2023 for Wind. Occlusion is an interpretation method that hides part of the input and sees how it impacts the CNN prediction. The higher the impact, the higher the hidden part importance           \cite{zeiler2013}. Power supply data is obtained from RTE for all France's regions (NUTS1)}
    \label{fig:cnn_interpretation}
\end{figure}

\noindent Although the increase in complexity between dimension reduction and spatial average approach did not lead to clear improvements in model performance for every model architecture, leveraging the entire weather maps with a more complex computer vision architecture such as a CNN clearly did. This phenomenom stems from the unsupervised nature of the PCA compared to the supervised CNN. In fact, the CNN is the best-performing model on the Wind dataset and the second-best on the Solar dataset. By utilizing our spatio-temporal weighting scheme, CNN has a better grasp of the trends in renewables implementation as highlighted in figure \ref{fig:cnn_interpretation}, and avoids extrapolation difficulties. Combined with the MLP results, it highlights the versatility and suitability of neural network-based models for predicting power production from renewable sources.\\

\noindent Tables \ref{tab:solar_benchmark_results} and \ref{tab:wind_benchmark_results} illustrate the challenges that tree-based models face with extrapolation. Without the detrending scheme, these models would not rank among the top three performers. Instead, neural networks would dominate the podium, with the rankings reflecting the increasing complexity of the modeling approaches. Specifically, as models incorporate more spatially explicit data, their performance improves, with vision models outperforming MLPs combined with PCA, which in turn surpass MLPs on time series. Therefore, we recommend that practitioners incorporate spatial information when designing forecasting models. \\

\noindent The work conducted on cross-validation procedures and HPO schemes allowed us to push state-of-the-art machine learning architecture at their best performance. However, such a study could be extended to include deep learning models such as CNN in order to improve their performance. As deep convolutional neural networks are already amongst the best models for both Solar and Wind, we did not pursue this path. However, it is worth mentioning that a systematic study would benefit deep learning models and strengthen their edge.\\

\section{Conclusion}

\noindent In this study, we built spatially explicit datasets for predicting solar and wind power supply based on weather data and power facilities locations and capacities over France, which are then used with other non-spatially explicit predictors, to predict daily solar and wind power generation at national scale over the last decade. Machine learning models with different architectures are tested, with specific tests to select the best calibration procedure that gives a generalization error as close as possible to the real error, obtained from model prediction on unseen data.\\

\noindent We conducted experiments by varying cross-validation methods and hyperparameter optimization algorithms to find out which settings is best suited to reach a realistic generalization error. We found that cross-validation methods that preserve the temporal structure of the dataset perform better, as expected for the problem of predicting time series with time-dependent data. The results also showed that some cross-validation and hyperparameter optimization methods tend to underestimate or overestimate the generalization error. This is a key insight to have, as it may explain the discrepancies between performance during model selection and performance on unseen data. We also showed that dataset's size impacts the generalization error estimate. Adding older and older data to the training set can enhance or worsen the precision of the cross-validation estimate depending on the method. However, we found that, as with any data-driven study, the results were sensitive to the data and the model at stake.\\

\noindent A benchmark of state-of-the-art architectures using three different modeling approaches dealing with spatially averaged or spatially explicit input data is also provided. We showed that the increase in renewable power production in recent years led to extrapolation difficulties on the test set, explaining the poor performance of some architectures, principally tree-based models. We tried to solve this extrapolation problem using linear trees or detrending. Although detrending always improved the test metrics of the models, linear trees were more sensitive to the data. We also showed that increasing model complexity by using dimension reduction techniques such as principal component analysis was not a silver bullet for prediction performance. However, computer vision architectures such as convolutional neural networks applied to spatially explicit input were efficient and amongst the best models for both Wind and Solar.

\noindent 
\paragraph{Competing Interests} The authors declare that they have no competing interests.
\newpage

\begin{appendix}
\section{Appendix A : Weather variables}\label{appendixA}
\begin{table}[ht!]
\resizebox{\textwidth}{!}{
\begin{tabular}{p{2.5cm} c c p{5cm} c}
\toprule
\textbf{Variable Full Name} & \textbf{Variable Abbreviation} & \textbf{Unit} & \textbf{Description} & \textbf{Sector}\\
\midrule
2 meter temperature & t2m & \SI{}{\kelvin} & Temperature of air at 2m above the surface & Solar, Wind\\

Surface solar radiation downwards & ssrd & \SI{}{\joule\per\meter\squared} & Amount of solar radiation (direct and diffuse) reaching a horizontal plane at the surface of the Earth & Solar\\

10 meters U wind component & u10 & \SI{}{\meter\per\second} & Northward component of the wind speed at 10m & Wind \\

10 meters V wind component & v10 & \SI{}{\meter\per\second} & Eastward component of the wind speed at 10m & Wind\\

100 meters U wind component & u100 & \SI{}{\meter\per\second} & Northward component of the wind speed at 100m & Wind \\

100 meters V wind component & v100 & \SI{}{\meter\per\second} & Eastward component of the wind speed at 100m & Wind \\

Instantaneous 10 meters wind gust & i10fg & \SI{}{\meter\per\second} & Maximum wind gust speed at 10m & Solar, Wind\\

Total Precipitation & tp & \SI{}{\meter} & Accumulated liquid and frozen water that falls to the Earth's surface & Wind\\

Evaporation & e & \SI{}{\meter} & Accumulated amount of water that has evaporated from the Earth's surface & Solar\\

Runoff & ro & \SI{}{\meter} & Water from rainfall, snow melt or deep soil that drains away over the surface or under the ground & Wind \\
\bottomrule
\end{tabular}}
\caption{Description of climate variables. \textit{Source: \href{https://confluence.ecmwf.int/display/CKB/ERA5\%3A+data+documentation}{ERA5 Documentation}}}
\label{tab:climate_variables}
\end{table}

\section{Appendix B : Metrics Definition}\label{appendixB}
\begin{equation}
  \mathrm{MAE} = \dfrac{1}{N}\sum_{i}^{N}|y_i -\hat{y}_i| \\
\end{equation}

\begin{equation}
  \mathrm{MAPE} = 100 \times \dfrac{1}{N}\sum_{i}^{N}|\dfrac{y_i -\hat{y}_i}{y_i}| \\
\end{equation}

\begin{equation}
  \mathrm{nRMSE} = 100 \times \dfrac{\sqrt{\dfrac{1}{N}\sum_{i}^{N}(y_i -\hat{y}_i)^2}}{y_{max} - y_{min}} \\
\end{equation}

\begin{equation}
  \mathrm{R2} = 1- \dfrac{\sum_{i}^{N}(y_i -\hat{y}_i)^2}{\sum_{i}^{N}(y_{i} - \overline{y})^2} \\
\end{equation}
with $y_{max}$, $y_{min}$, and $\overline{y}$ the maximum, minimum, and the average of the true target $y$.

\newpage
\section{Appendix C : Cross-Validation Experiment results for Solar}\label{appendixC}

\subsection{Boosting}
\begin{figure}[ht!]
    \centering
    \includegraphics[width=0.8\textwidth]{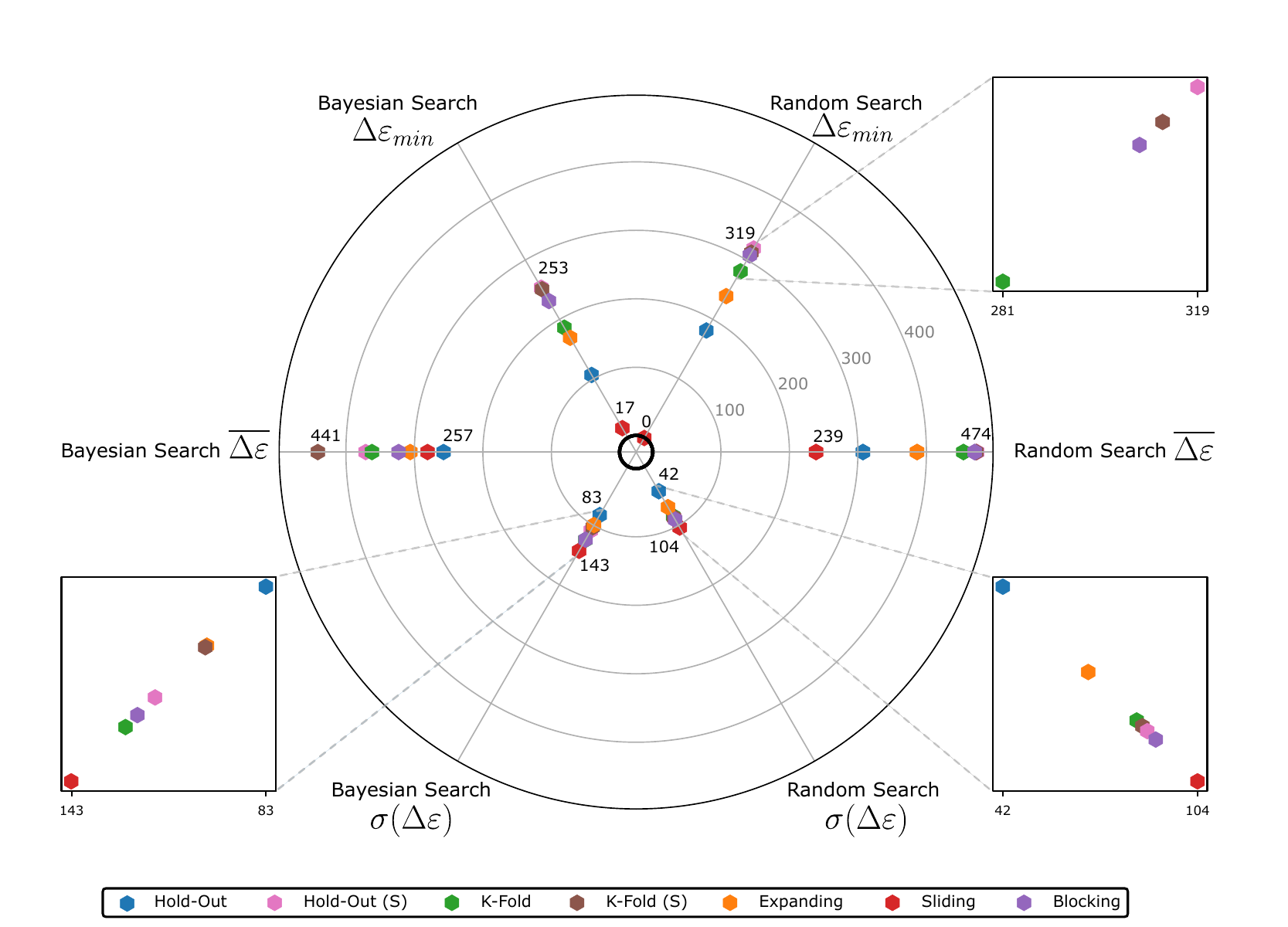}
    \caption{Results of different cross-validation techniques for boosted trees on Solar. Only the worst and best values for each axis are printed. The (S) indicates the shuffling variant of the method}
    \label{fig:cv_solar_gb}
\end{figure}

\begin{figure}[ht!]
    \centering
    \includegraphics[width=0.8\textwidth]{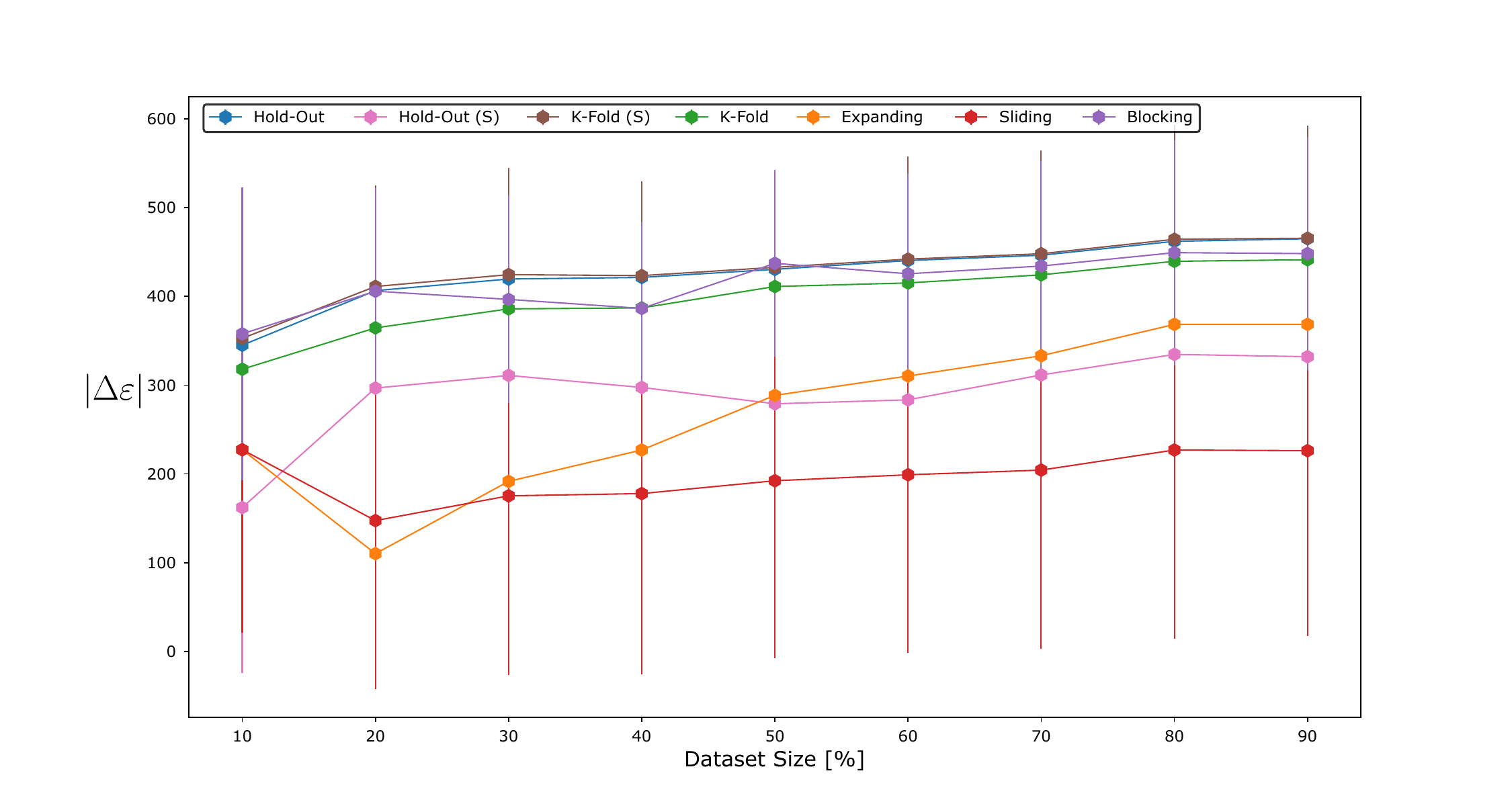}
    \caption{Robustness of cross-validation procedure regarding dataset size for boosted tress on Solar. The marker indicates the average $|\Delta \varepsilon|$ while the error bars display the standard deviation. The (S) indicates the shuffling variant of the method}
    \label{fig:cv_solar_gb_size}
\end{figure}

\newpage
\subsection{Feed-Forward Neural Network (MLP)}
\begin{figure}[ht!]
    \centering
    \includegraphics[width=0.8\textwidth]{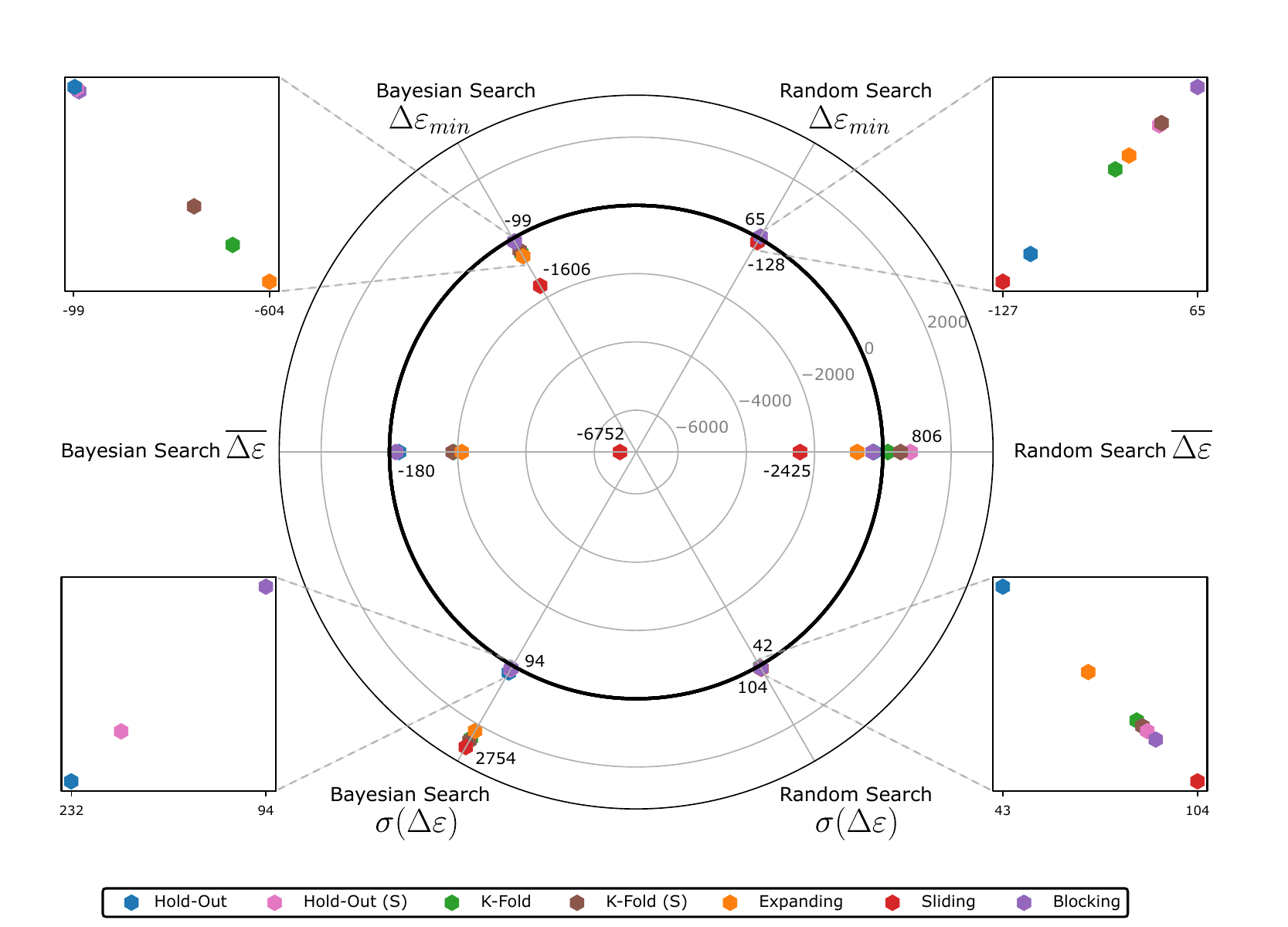}
    \caption{Results of different cross-validation techniques for feed-forward neural network on Solar. Only the worst and best values for each axis are printed. The (S) indicates the shuffling variant of the method}
    \label{fig:cv_solar_nn}
\end{figure}

\begin{figure}[ht!]
    \centering
    \includegraphics[width=0.8\textwidth]{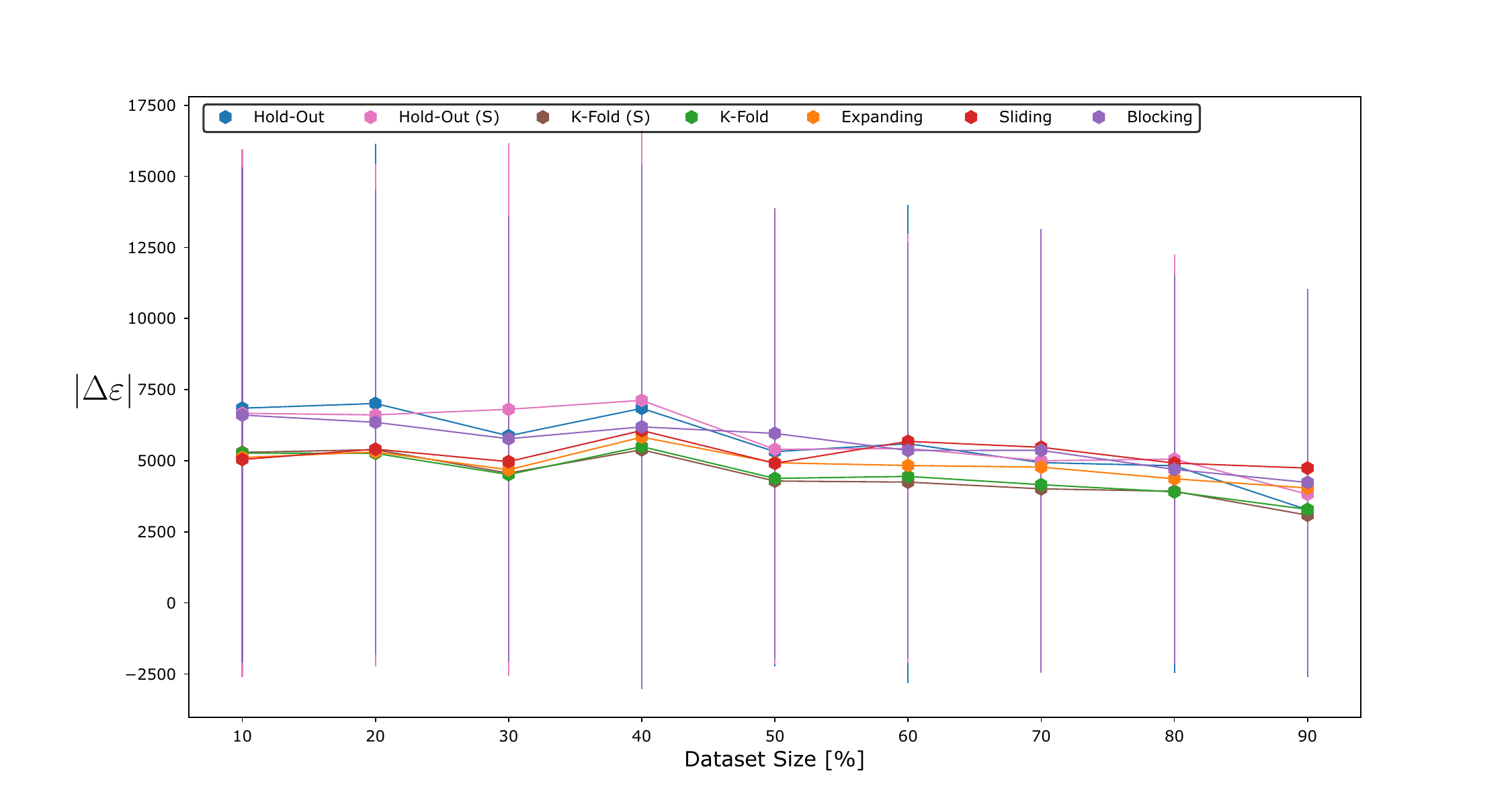}
    \caption{Robustness of cross-validation procedure regarding dataset size for feed-forward neural network on Solar. The marker indicates the average $|\Delta \varepsilon|$ while the error bars display the standard deviation. The (S) indicates the shuffling variant of the method}
    \label{fig:cv_solar_nn_size}
\end{figure}

\section{Appendix D : Cross-Validation Experiment results for Wind}\label{appendixD}

\subsection{Random Forest}
\begin{figure}[ht!]
    \centering
    \includegraphics[width=0.8\textwidth]{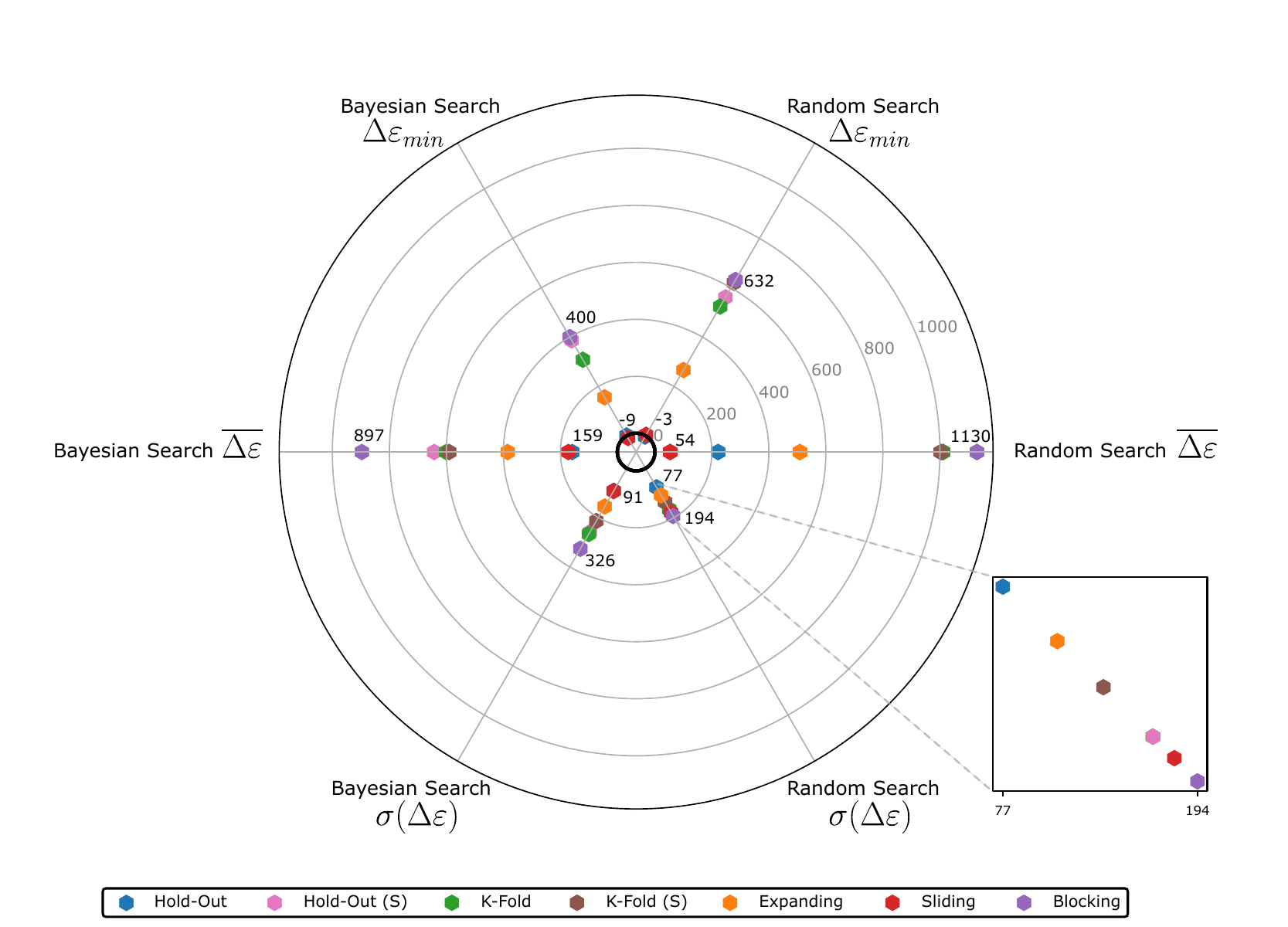}
    \caption{Results of different cross-validation techniques for random forest on Wind. Only the worst and best values for each axis are printed. The (S) indicates the shuffling variant of the method}
    \label{fig:cv_wind_rf}
\end{figure}

\begin{figure}[ht!]
    \centering
    \includegraphics[width=0.8\textwidth]{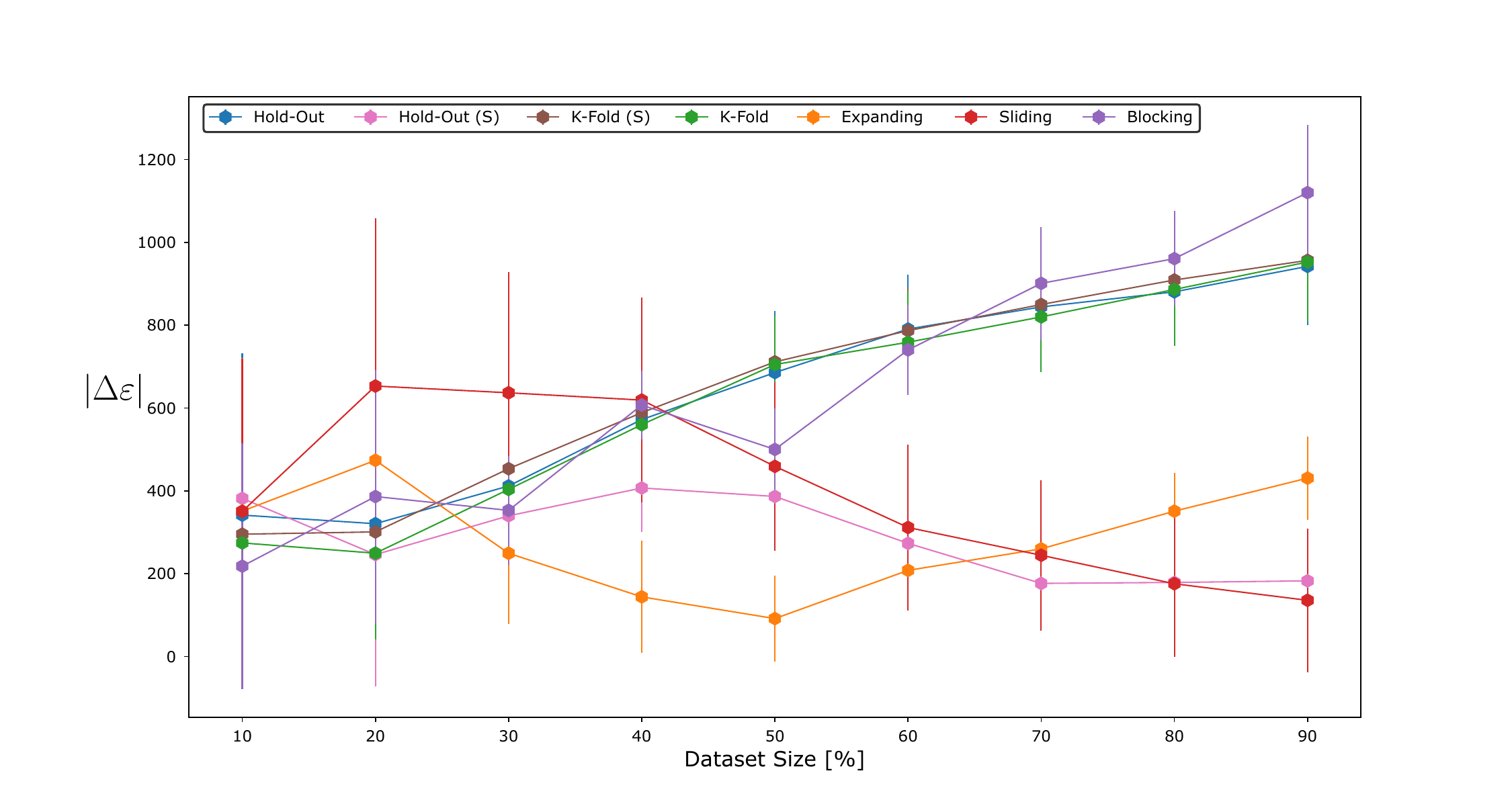}
    \caption{Robustness of cross-validation procedure regarding dataset size for random forest on Wind. The marker indicates the average $|\Delta \varepsilon|$ while the error bars display the standard deviation. The (S) indicates the shuffling variant of the method}
    \label{fig:cv_wind_rf_size}
\end{figure}

\subsection{Boosting}
\begin{figure}[ht!]
    \centering
    \includegraphics[width=0.8\textwidth]{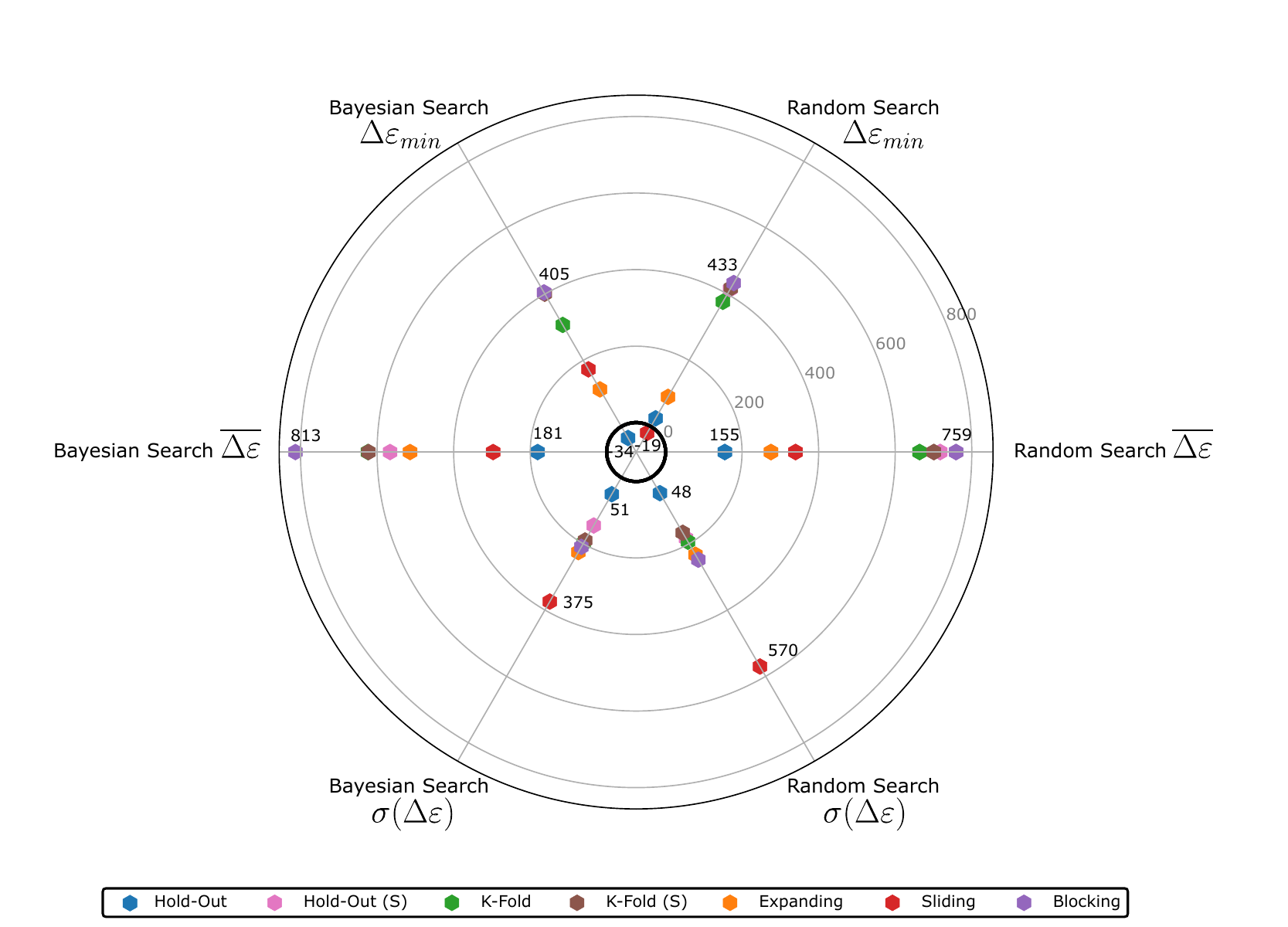}
    \caption{Results of different cross-validation techniques for boosted trees on Wind. Only the worst and best values for each axis are printed. The (S) indicates the shuffling variant of the method}
    \label{fig:cv_wind_gb}
\end{figure}

\begin{figure}[ht!]
    \centering
    \includegraphics[width=0.8\textwidth]{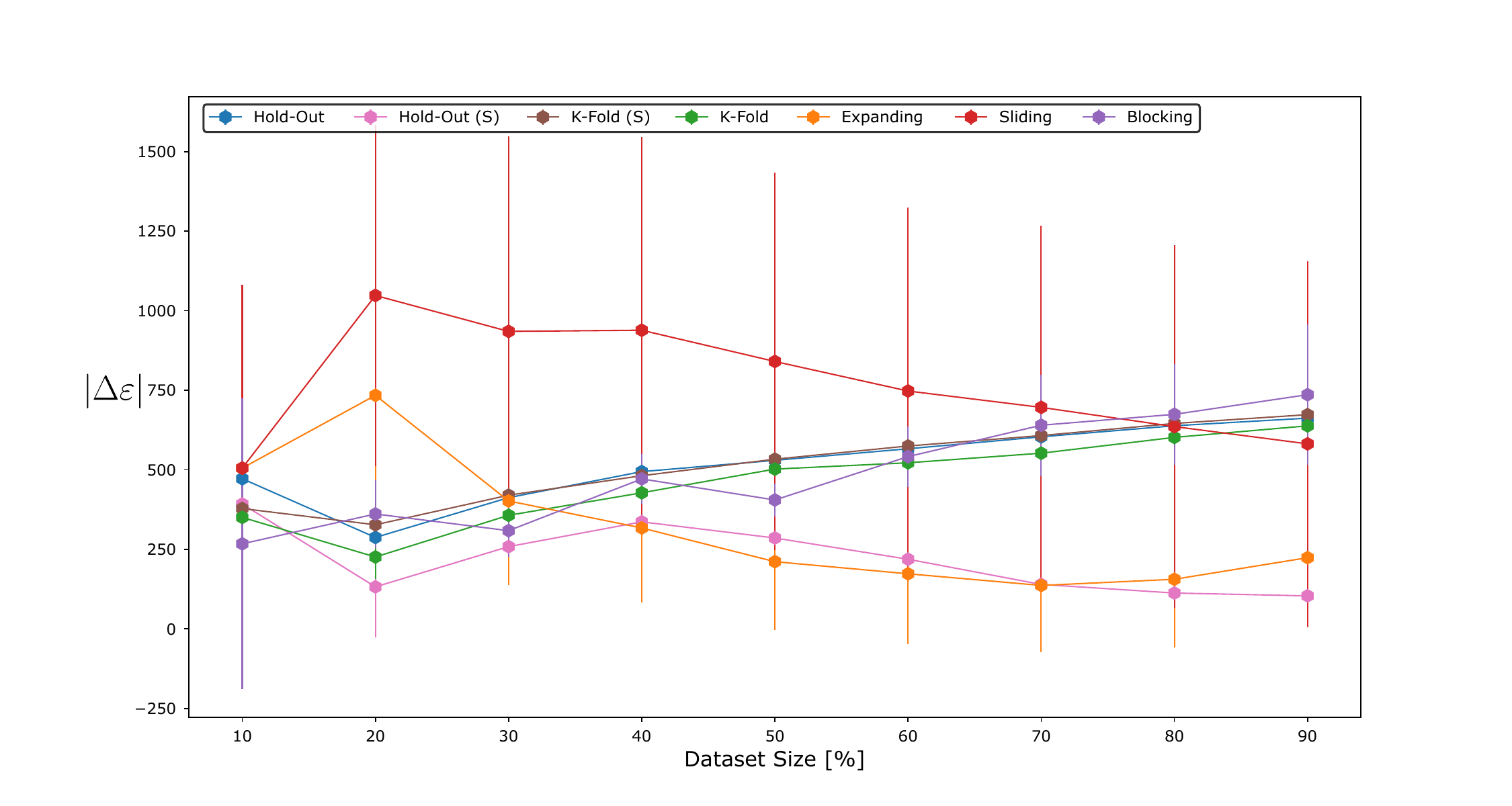}
    \caption{Robustness of cross-validation procedure regarding dataset size for boosted trees on Wind. The marker indicates the average $|\Delta \varepsilon|$ while the error bars display the standard deviation. The (S) indicates the shuffling variant of the method}
    \label{fig:cv_wind_gb_size}
\end{figure}

\newpage 
\subsection{Feed-Forward Neural Network (MLP)}
\begin{figure}[ht!]
    \centering
    \includegraphics[width=0.8\textwidth]{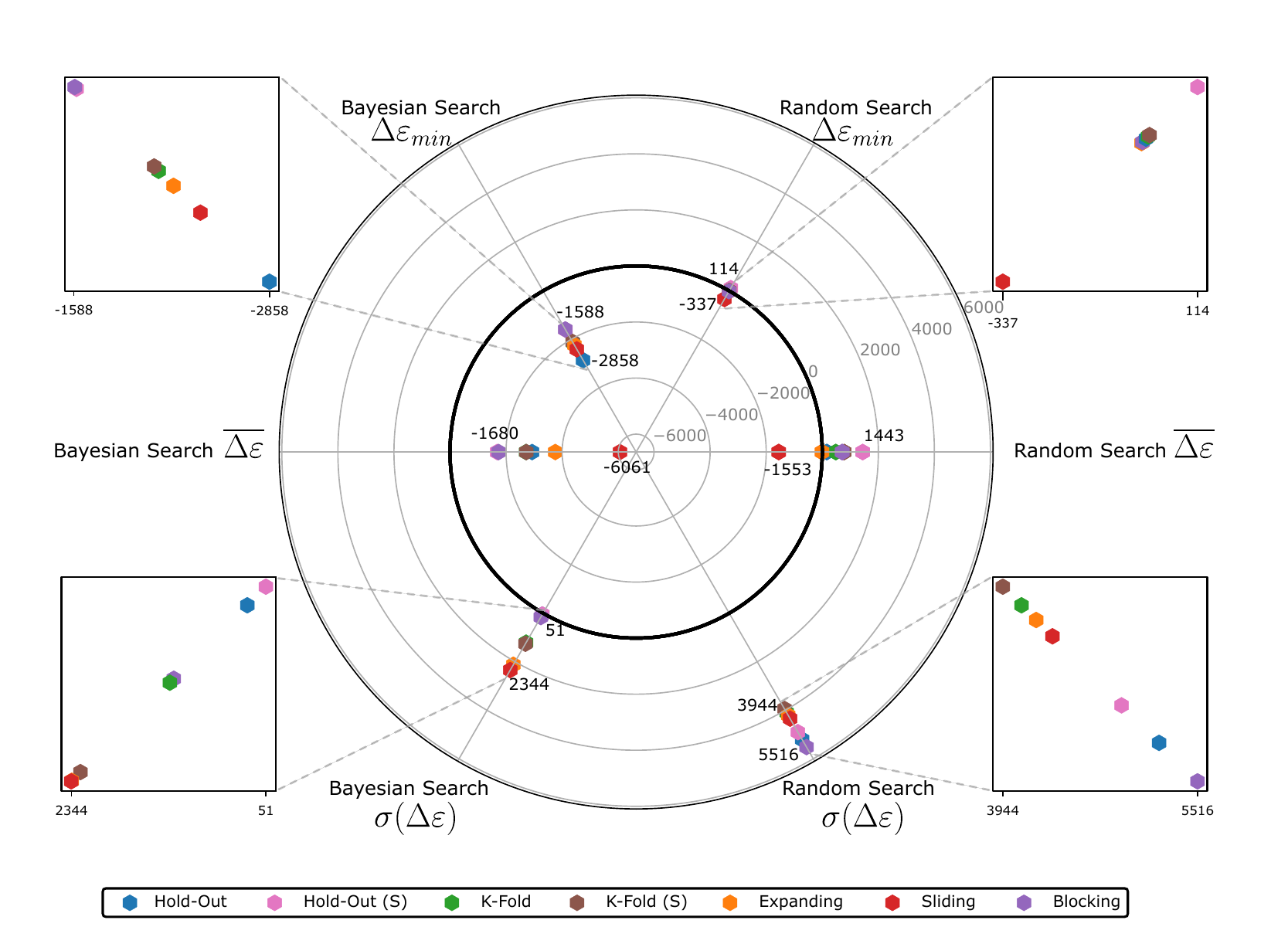}
    \caption{Results of different cross-validation techniques for feed-forward neural network on Wind. Only the worst and best values for each axis are printed. The (S) indicates the shuffling variant of the method}
    \label{fig:cv_wind_nn}
\end{figure}

\begin{figure}[ht!]
    \centering
    \includegraphics[width=0.8\textwidth]{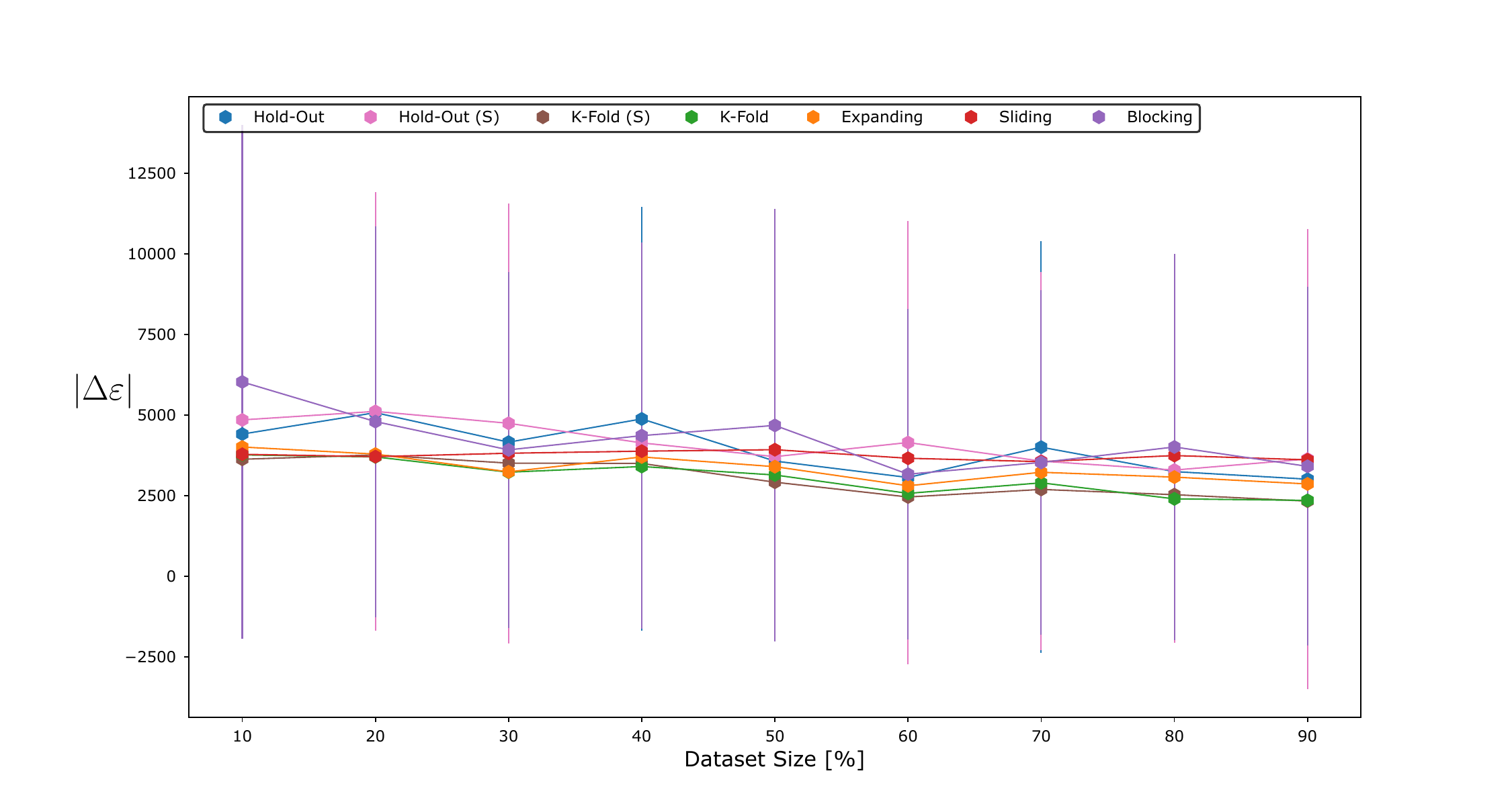}
    \caption{Robustness of cross-validation procedure regarding dataset size for feed-forward neural network on Wind. The marker indicates the average $|\Delta \varepsilon|$ while the error bars display the standard deviation. The (S) indicates the shuffling variant of the method}
    \label{fig:cv_wind_nn_size}
\end{figure}

\newpage
\section{Appendix E : Benchmark results for Wind}\label{appendixE}
\begin{table}[ht]
\resizebox{\textwidth}{!}{
\begin{tabular}{l|l|c|ll|ll|ll|ll|ll}
\multicolumn{3}{c}{Metrics} & \multicolumn{2}{c}{\textbf{MAE}} & \multicolumn{2}{c}{\textbf{MAPE (\%)}} & \multicolumn{2}{c}{\textbf{RMSE}} & \multicolumn{2}{c}{\textbf{nRMSE (\%)}} & \multicolumn{2}{c}{\textbf{R2}} \\
\toprule
\textbf{Approach} & \textbf{Model} & \textbf{Detrend} & Train & Test & Train & Test & Train & Test & Train & Test & Train & Test \\
\midrule
\midrule
Average & Linear Regression & & 313 & 834 & 15.8 & 19.5 & 424 & 1104 & 2.73 & 7.24 & 0.97 & 0.90\\
Average & Random Forest & & 99.2 & 1180 & 4.27 & 25.6 & 140 & 1464 & 0.90 & 9.60 & 1.0 & 0.83 \\
Average & Random Forest & \checkmark & 112 & 483 \bronzemedal & 4.80 & 11.3 \bronzemedal & 159 & 650 & 1.03 & 4.26 & 1.0 & 0.97 \\
Average & Linear Forest & & 260 & 707 & 12.2 & 19.8 & 354 & 967 & 2.28 & 6.35 & 0.98 & 0.93 \\
Average & Tree Boosting & & 92.6 & 901 & 2.90 & 19.1 & 145 & 1173 & 0.93 & 7.70 & 1.0 & 0.89 \\
Average & Tree Boosting & \checkmark & 181 & 496 & 8.04 & 11.9 & 250 & 650\bronzemedal & 1.61 & 4.26 \bronzemedal & 0.99 & 0.97 \bronzemedal\\
Average & Linear Tree Boosting & & 361 & 748 & 17.2 & 19.8 &484 & 1042 & 3.12& 6.84 &0.96 & 0.92 \\
Average & GAM & & 243 & 628 & 11.8 & 17.4 & 332 & 807 & 2.14 & 5.29 & 0.98 & 0.95 \\
Average & MLP & & 267 & 438 \silvermedal & 10.8 & 9.56 \silvermedal & 383 & 608 \silvermedal & 2.47 & 3.99 \silvermedal& 0.98 & 0.97 \silvermedal\\
PCA & Linear Regression & & 484 & 799 & 22.6 & 19.1 & 654 & 1089 & 4.21 & 7.14 & 0.93 & 0.91 \\
PCA & Random Forest & & 177 & 1168 & 8.27 & 26.1 & 252 & 1500 & 1.63 & 9.84 & 0.99 & 0.82 \\
PCA & Linear Forest & & 159 & 1233 & 7.24 & 30.3 & 240 & 1551 & 1.55 & 10.2 & 0.99 & 0.81 \\
PCA & Tree Boosting & & 156& 803 & 8.14 & 17.5 & 204 & 1057 & 1.32 & 6.93 &0.99 &0.91 \\
PCA & Linear Tree Boosting & & 423 & 801 & 19.7 & 19.0 & 574 & 1066 & 3.70 & 6.99 & 0.94 & 0.91 \\
PCA & GAM & & 328 & 750 & 16.3 & 20.2 & 438 & 929 & 2.82 & 6.10 & 0.97 & 0.93\\
PCA & MLP & & 283 & 508 & 12.6& 11.5 & 373 & 693& 2.40 & 4.54& 0.98 &0.96 \\
Vision & CNN & & 240& 417 \goldmedal & 9.98& 9.12 \goldmedal & 340 & 575 \goldmedal & 2.19 & 3.77 \goldmedal & 0.98 & 0.97 \goldmedal\\
\bottomrule
\end{tabular}}
\caption{Benchmark results for different models using 3 different modeling approaches on the Wind dataset. Medals indicate the top three best-performing models on the test set for each metric}
\label{tab:wind_benchmark_results}
\end{table}
\end{appendix}

\newpage
\bibliography{references}  %%% Uncomment this line and comment out the ``thebibliography'' section below to use the external .bib file (using bibtex) .

\begin{thebibliography}{72}
\providecommand{\natexlab}[1]{#1}
\providecommand{\url}[1]{\texttt{#1}}
\expandafter\ifx\csname urlstyle\endcsname\relax
  \providecommand{\doi}[1]{doi: #1}\else
  \providecommand{\doi}{doi: \begingroup \urlstyle{rm}\Url}\fi

\bibitem[Abdul~Baseer et~al.(2023)Abdul~Baseer, Almunif, Alsaduni, and Tazeen]{abdul_baseer_electrical_2023}
Mohammad Abdul~Baseer, Anas Almunif, Ibrahim Alsaduni, and Nazia Tazeen.
\newblock Electrical power generation forecasting from renewable energy systems using artificial intelligence techniques.
\newblock \emph{Energies}, 16:\penalty0 6414, 2023.
\newblock ISSN 1996-1073.
\newblock \doi{10.3390/en16186414}.
\newblock URL \url{https://www.mdpi.com/1996-1073/16/18/6414}.

\bibitem[Ahmad and Hossain(2020)]{ahmad_maximizing_2020}
Shahryar~Khalique Ahmad and Faisal Hossain.
\newblock Maximizing energy production from hydropower dams using short-term weather forecasts.
\newblock \emph{Renewable Energy}, 146:\penalty0 1560--1577, 2020.
\newblock ISSN 09601481.
\newblock \doi{10.1016/j.renene.2019.07.126}.
\newblock URL \url{https://linkinghub.elsevier.com/retrieve/pii/S0960148119311462}.

\bibitem[Alcañiz et~al.(2023)Alcañiz, Lindfors, Zeman, Ziar, and Isabella]{alcaniz_effect_2023}
Alba Alcañiz, Anders~V. Lindfors, Miro Zeman, Hesan Ziar, and Olindo Isabella.
\newblock Effect of climate on photovoltaic yield prediction using machine learning models.
\newblock \emph{Global Challenges}, 7\penalty0 (1):\penalty0 2200166, 2023.
\newblock ISSN 2056-6646, 2056-6646.
\newblock \doi{10.1002/gch2.202200166}.
\newblock URL \url{https://onlinelibrary.wiley.com/doi/10.1002/gch2.202200166}.

\bibitem[Arlot and Celisse(2009)]{arlot_2009}
Sylvain Arlot and Alain Celisse.
\newblock A survey of cross validation procedures for model selection.
\newblock \emph{Statistics Surveys}, 4, 07 2009.
\newblock \doi{10.1214/09-SS054}.

\bibitem[Bellinguer et~al.(2020)Bellinguer, Mahler, Camal, and Kariniotakis]{bellinguer_2020}
Kevin Bellinguer, Valentin Mahler, Simon Camal, and Georges Kariniotakis.
\newblock {Probabilistic Forecasting of Regional Wind Power Generation for the EEM20 Competition: a Physics-oriented Machine Learning Approach}.
\newblock In \emph{{17th European Energy Market Conference, EEM 2020}}, Stockholm, Sweden, September 2020. {KTH, IEEE}.
\newblock URL \url{https://hal.science/hal-02952589}.

\bibitem[Bergmeir and Benítez(2012)]{bergmeir_2012}
Christoph Bergmeir and José~M. Benítez.
\newblock On the use of cross-validation for time series predictor evaluation.
\newblock \emph{Information Sciences}, 191:\penalty0 192--213, 2012.
\newblock ISSN 0020-0255.
\newblock \doi{https://doi.org/10.1016/j.ins.2011.12.028}.
\newblock URL \url{https://www.sciencedirect.com/science/article/pii/S0020025511006773}.
\newblock Data Mining for Software Trustworthiness.

\bibitem[Bergstra et~al.(2011)Bergstra, Bardenet, Bengio, and Kegl]{bergstra_algorithms_nodate}
James Bergstra, Remi Bardenet, Yoshua Bengio, and Balazs Kegl.
\newblock Algorithms for hyper-parameter optimization.
\newblock In \emph{Neural Information Processing Systems Proceedings (NeurIPS)}, 2011.

\bibitem[Biber et~al.(2022)Biber, Felder, Wieland, and Spliethoff]{bieber_2022}
Albert Biber, Mine Felder, Christoph Wieland, and Hartmut Spliethoff.
\newblock Negative price spiral caused by renewables ? {E}lectricity price prediction on the german market for 2030.
\newblock \emph{The Electricity Journal}, 35\penalty0 (8):\penalty0 107188, 2022.
\newblock ISSN 1040-6190.
\newblock \doi{https://doi.org/10.1016/j.tej.2022.107188}.
\newblock URL \url{https://www.sciencedirect.com/science/article/pii/S1040619022001142}.

\bibitem[Bilendo et~al.(2023)Bilendo, Meyer, Badihi, Lu, Cambron, and Jiang]{bilendo_2023}
Francisco Bilendo, Angela Meyer, Hamed Badihi, Ningyun Lu, Philippe Cambron, and Bin Jiang.
\newblock Applications and modeling techniques of wind turbine power curve for wind farms: A review.
\newblock \emph{Energies}, 16\penalty0 (1), 2023.
\newblock ISSN 1996-1073.
\newblock \doi{10.3390/en16010180}.
\newblock URL \url{https://www.mdpi.com/1996-1073/16/1/180}.

\bibitem[Bischl et~al.(2023)Bischl, Binder, Lang, Pielok, Richter, Coors, Thomas, Ullmann, Becker, Boulesteix, Deng, and Lindauer]{bischl_hyperparameter_2023}
Bernd Bischl, Martin Binder, Michel Lang, Tobias Pielok, Jakob Richter, Stefan Coors, Janek Thomas, Theresa Ullmann, Marc Becker, Anne‐Laure Boulesteix, Difan Deng, and Marius Lindauer.
\newblock Hyperparameter optimization: Foundations, algorithms, best practices, and open challenges.
\newblock \emph{{WIREs} Data Mining and Knowledge Discovery}, 13\penalty0 (2):\penalty0 e1484, 2023.
\newblock ISSN 1942-4787, 1942-4795.
\newblock \doi{10.1002/widm.1484}.
\newblock URL \url{https://wires.onlinelibrary.wiley.com/doi/10.1002/widm.1484}.

\bibitem[{British Petroleum (BP)}(2024)]{bp_energy_outlook}
{British Petroleum (BP)}.
\newblock {Energy Outlook 2024: Exploring the key trends and uncertainties surrounding the energy transition}.
\newblock \url{https://www.bp.com/en/global/corporate/energy-economics/energy-outlook.html}, 2024.
\newblock Accessed: August 26, 2024.

\bibitem[Castillo-Rojas et~al.(2023)Castillo-Rojas, Medina~Quispe, and Hernández]{castillo_2023}
Wilson Castillo-Rojas, Fernando Medina~Quispe, and César Hernández.
\newblock Photovoltaic energy forecast using weather data through a hybrid model of recurrent and shallow neural networks.
\newblock \emph{Energies}, 16\penalty0 (13), 2023.
\newblock ISSN 1996-1073.
\newblock \doi{10.3390/en16135093}.
\newblock URL \url{https://www.mdpi.com/1996-1073/16/13/5093}.

\bibitem[Cerqueira et~al.(2019)Cerqueira, Torgo, and Mozetic]{cerqueira_2019}
Vitor Cerqueira, Luís Torgo, and Igor Mozetic.
\newblock Evaluating time series forecasting models: An empirical study on performance estimation methods.
\newblock \emph{Machine Learning}, 05 2019.
\newblock \doi{10.48550/arXiv.1905.11744}.

\bibitem[Chatfield(1986)]{chatfield_1986}
Chris Chatfield.
\newblock Comparative models for electrical load forecasting.
\newblock \emph{Journal of the Royal Statistical Society. Series A (General)}, 149\penalty0 (3):\penalty0 272--272, 1986.
\newblock ISSN 00359238, 23972327.
\newblock URL \url{http://www.jstor.org/stable/2981560}.

\bibitem[Chen et~al.(2023)Chen, Hu, Wang, Wang, and Zhu]{gang_2023}
Gang Chen, Qingchang Hu, Jin Wang, Xu~Wang, and Yuyu Zhu.
\newblock Machine-learning-based electric power forecasting.
\newblock \emph{Sustainability}, 15\penalty0 (14), 2023.
\newblock ISSN 2071-1050.
\newblock \doi{10.3390/su151411299}.
\newblock URL \url{https://www.mdpi.com/2071-1050/15/14/11299}.

\bibitem[Chen and Folly(2018)]{chen_wind_2018}
Q.~Chen and K.A. Folly.
\newblock Wind power forecasting.
\newblock \emph{{IFAC}-{PapersOnLine}}, 51\penalty0 (28):\penalty0 414--419, 2018.
\newblock ISSN 24058963.
\newblock \doi{10.1016/j.ifacol.2018.11.738}.
\newblock URL \url{https://linkinghub.elsevier.com/retrieve/pii/S240589631833458X}.

\bibitem[Condemi et~al.(2021)Condemi, Casillas-Pérez, Mastroeni, Jiménez-Fernández, and Salcedo-Sanz]{condemi_hydro-power_2021}
C.~Condemi, D.~Casillas-Pérez, L.~Mastroeni, S.~Jiménez-Fernández, and S.~Salcedo-Sanz.
\newblock Hydro-power production capacity prediction based on machine learning regression techniques.
\newblock \emph{Knowledge-Based Systems}, 222:\penalty0 107012, 2021.
\newblock ISSN 09507051.
\newblock \doi{10.1016/j.knosys.2021.107012}.
\newblock URL \url{https://linkinghub.elsevier.com/retrieve/pii/S0950705121002756}.

\bibitem[Couto and Estanqueiro(2022)]{couto_enhancing_2022}
António Couto and Ana Estanqueiro.
\newblock Enhancing wind power forecast accuracy using the weather research and forecasting numerical model-based features and artificial neuronal networks.
\newblock \emph{Renewable Energy}, 201:\penalty0 1076--1085, 2022.
\newblock ISSN 09601481.
\newblock \doi{10.1016/j.renene.2022.11.022}.
\newblock URL \url{https://linkinghub.elsevier.com/retrieve/pii/S096014812201655X}.

\bibitem[De~Giorgi et~al.(2014)De~Giorgi, Congedo, and Malvoni]{de_giorgi_photovoltaic_2014}
Maria~Grazia De~Giorgi, Paolo~Maria Congedo, and Maria Malvoni.
\newblock Photovoltaic power forecasting using statistical methods: impact of weather data.
\newblock \emph{{IET} Science, Measurement \& Technology}, 8\penalty0 (3):\penalty0 90--97, 2014.
\newblock ISSN 1751-8830, 1751-8830.
\newblock \doi{10.1049/iet-smt.2013.0135}.
\newblock URL \url{https://onlinelibrary.wiley.com/doi/10.1049/iet-smt.2013.0135}.

\bibitem[De~Vita et~al.(2018)De~Vita, Capros, Evangelopoulou, Kannavou, Siskos, and Zazias]{devita_2018}
Alessia De~Vita, Pantelis Capros, Stavroula Evangelopoulou, Maria Kannavou, Pelopidas Siskos, and Georgios Zazias.
\newblock Sectoral integration : Long-term perspective in the {EU} energy system, 02 2018.

\bibitem[Dolara et~al.(2015)Dolara, Leva, and Manzolini]{dolara_2015}
Alberto Dolara, Sonia Leva, and Giampaolo Manzolini.
\newblock Comparison of different physical models for pv power output prediction.
\newblock \emph{Solar Energy}, 119:\penalty0 83--99, 2015.
\newblock ISSN 0038-092X.
\newblock \doi{https://doi.org/10.1016/j.solener.2015.06.017}.
\newblock URL \url{https://www.sciencedirect.com/science/article/pii/S0038092X15003254}.

\bibitem[Elsaraiti and Merabet(2022)]{elsaraiti_solar_2022}
Meftah Elsaraiti and Adel Merabet.
\newblock Solar power forecasting using deep learning techniques.
\newblock \emph{{IEEE} Access}, 10:\penalty0 31692--31698, 2022.
\newblock ISSN 2169-3536.
\newblock \doi{10.1109/ACCESS.2022.3160484}.
\newblock URL \url{https://ieeexplore.ieee.org/document/9737470/}.

\bibitem[Engeland et~al.(2017)Engeland, Borga, Creutin, François, Ramos, and Vidal]{engeland_2017}
Kolbjørn Engeland, Marco Borga, Jean-Dominique Creutin, Baptiste François, Maria-Helena Ramos, and Jean-Philippe Vidal.
\newblock Space-time variability of climate variables and intermittent renewable electricity production – a review.
\newblock \emph{Renewable and Sustainable Energy Reviews}, 79:\penalty0 600--617, 2017.
\newblock ISSN 1364-0321.
\newblock \doi{https://doi.org/10.1016/j.rser.2017.05.046}.
\newblock URL \url{https://www.sciencedirect.com/science/article/pii/S1364032117306822}.

\bibitem[{European Comission}(2019)]{european_green_deal}
{European Comission}.
\newblock {The European Green Deal: Striving to be the first climate-neutral continent}.
\newblock \url{https://commission.europa.eu/strategy-and-policy/priorities-2019-2024/european-green-deal_en}, 2019.
\newblock Accessed: August 26, 2024.

\bibitem[Gaillard et~al.(2016)Gaillard, Goude, and Nedellec]{gaillard_2016}
Pierre Gaillard, Yannig Goude, and Raphaël Nedellec.
\newblock Additive models and robust aggregation for {GEFC}om2014 probabilistic electric load and electricity price forecasting.
\newblock \emph{International Journal of Forecasting}, 32\penalty0 (3):\penalty0 1038--1050, 2016.
\newblock ISSN 0169-2070.
\newblock \doi{https://doi.org/10.1016/j.ijforecast.2015.12.001}.
\newblock URL \url{https://www.sciencedirect.com/science/article/pii/S0169207015001545}.

\bibitem[Gama and Brazdil(1999)]{gama_1999}
João Gama and Pavel Brazdil.
\newblock Linear tree.
\newblock \emph{Intelligent Data Analysis}, 3\penalty0 (1):\penalty0 1--22, 1999.
\newblock ISSN 1088-467X.
\newblock \doi{https://doi.org/10.1016/S1088-467X(99)00002-5}.
\newblock URL \url{https://www.sciencedirect.com/science/article/pii/S1088467X99000025}.

\bibitem[Gijón et~al.(2023)Gijón, Pujana-Goitia, Perea, Molina-Solana, and Gómez-Romero]{gijon_prediction_2023}
Alfonso Gijón, Ainhoa Pujana-Goitia, Eugenio Perea, Miguel Molina-Solana, and Juan Gómez-Romero.
\newblock Prediction of wind turbines power with physics-informed neural networks and evidential uncertainty quantification, 2023.
\newblock URL \url{http://arxiv.org/abs/2307.14675}.

\bibitem[Goude et~al.(2014)Goude, Nedellec, and Kong]{goude_2016}
Yannig Goude, Raphael Nedellec, and Nicolas Kong.
\newblock Local short and middle term electricity load forecasting with semi-parametric additive models.
\newblock \emph{IEEE Transactions on Smart Grid}, 5\penalty0 (1):\penalty0 440--446, 2014.
\newblock \doi{10.1109/TSG.2013.2278425}.

\bibitem[Hengl et~al.(2018)Hengl, Nussbaum, Wright, Heuvelink, and Graeler]{hengl_2018}
Tomislav Hengl, Madlene Nussbaum, Marvin Wright, Gerard Heuvelink, and Benedikt Graeler.
\newblock Random forest as a generic framework for predictive modeling of spatial and spatio-temporal variables.
\newblock \emph{PeerJ}, 05 2018.
\newblock \doi{10.7287/peerj.preprints.26693v2}.

\bibitem[Hersbach et~al.(2020)Hersbach, Bell, Berrisford, Hirahara, Horányi, Muñoz‐Sabater, Nicolas, Peubey, Radu, Schepers, Simmons, Soci, Abdalla, Abellan, Balsamo, Bechtold, Biavati, Bidlot, Bonavita, De~Chiara, Dahlgren, Dee, Diamantakis, Dragani, Flemming, Forbes, Fuentes, Geer, Haimberger, Healy, Hogan, Hólm, Janisková, Keeley, Laloyaux, Lopez, Lupu, Radnoti, De~Rosnay, Rozum, Vamborg, Villaume, and Thépaut]{hersbach_era5_2020}
Hans Hersbach, Bill Bell, Paul Berrisford, Shoji Hirahara, András Horányi, Joaquín Muñoz‐Sabater, Julien Nicolas, Carole Peubey, Raluca Radu, Dinand Schepers, Adrian Simmons, Cornel Soci, Saleh Abdalla, Xavier Abellan, Gianpaolo Balsamo, Peter Bechtold, Gionata Biavati, Jean Bidlot, Massimo Bonavita, Giovanna De~Chiara, Per Dahlgren, Dick Dee, Michail Diamantakis, Rossana Dragani, Johannes Flemming, Richard Forbes, Manuel Fuentes, Alan Geer, Leo Haimberger, Sean Healy, Robin~J. Hogan, Elías Hólm, Marta Janisková, Sarah Keeley, Patrick Laloyaux, Philippe Lopez, Cristina Lupu, Gabor Radnoti, Patricia De~Rosnay, Iryna Rozum, Freja Vamborg, Sebastien Villaume, and Jean‐Noël Thépaut.
\newblock The {ERA}5 global reanalysis.
\newblock \emph{Quarterly Journal of the Royal Meteorological Society}, 146\penalty0 (730):\penalty0 1999--2049, 2020.
\newblock ISSN 0035-9009, 1477-870X.
\newblock \doi{10.1002/qj.3803}.
\newblock URL \url{https://rmets.onlinelibrary.wiley.com/doi/10.1002/qj.3803}.

\bibitem[Hyndman and Athanasopoulos(2018)]{hyndman}
{Robin John} Hyndman and George Athanasopoulos.
\newblock \emph{Forecasting: Principles and Practice}.
\newblock OTexts, Australia, 2nd edition, 2018.

\bibitem[Iheanetu(2022)]{iheanetu_solar_2022}
Kelachukwu~J. Iheanetu.
\newblock Solar photovoltaic power forecasting: A review.
\newblock \emph{Sustainability}, 14\penalty0 (24):\penalty0 17005, 2022.
\newblock ISSN 2071-1050.
\newblock \doi{10.3390/su142417005}.
\newblock URL \url{https://www.mdpi.com/2071-1050/14/24/17005}.

\bibitem[Ilic et~al.(2021)Ilic, Görgülü, Cevik, and Baydoğan]{ilic_2021}
Igor Ilic, Berk Görgülü, Mucahit Cevik, and Mustafa~Gökçe Baydoğan.
\newblock Explainable boosted linear regression for time series forecasting.
\newblock \emph{Pattern Recognition}, 120:\penalty0 108144, 2021.
\newblock ISSN 0031-3203.
\newblock \doi{https://doi.org/10.1016/j.patcog.2021.108144}.
\newblock URL \url{https://www.sciencedirect.com/science/article/pii/S0031320321003319}.

\bibitem[{International Renewable Energy Agency (IRENA)}(2020{\natexlab{a}})]{irena_innovation_brief}
{International Renewable Energy Agency (IRENA)}.
\newblock Advanced forecasting of variable renewable power generation: Innovation landscape brief, 2020{\natexlab{a}}.

\bibitem[{International Renewable Energy Agency (IRENA)}(2020{\natexlab{b}})]{irena_outlook}
{International Renewable Energy Agency (IRENA)}.
\newblock {Renewable Energy Prospects for Central and South-Eastern Europe Energy Connectivity (CESEC)}.
\newblock \url{https://www.irena.org/Publications/2020/Oct/Renewable-Energy-Prospects-for-Central-and-South-Eastern-Europe-Energy-Connectivity-CESEC}, 2020{\natexlab{b}}.
\newblock Accessed: August 26, 2024.

\bibitem[Keisler and Naour(2024)]{keisler_winddragon_2024}
Julie Keisler and Etienne~Le Naour.
\newblock {WINDDRAGON}: {Enhancing Wind Power Forecasting With Automated Deep Learning}.
\newblock \emph{Tackling Climate Change with Machine Learning, ICLR 2024}, 2024.

\bibitem[Kim et~al.(2017)Kim, Kim, Yoo, Lee, and Kim]{kim_daily_2017}
Jae‐Gon Kim, Dong‐Hyuk Kim, Woo‐Sik Yoo, Joung‐Yun Lee, and Yong~Bae Kim.
\newblock Daily prediction of solar power generation based on weather forecast information in korea.
\newblock \emph{{IET} Renewable Power Generation}, 11\penalty0 (10):\penalty0 1268--1273, 2017.
\newblock ISSN 1752-1416, 1752-1424.
\newblock \doi{10.1049/iet-rpg.2016.0698}.
\newblock URL \url{https://onlinelibrary.wiley.com/doi/10.1049/iet-rpg.2016.0698}.

\bibitem[Kim et~al.(2019)Kim, Jung, and Sim]{kim_two-step_2019}
Seul-Gi Kim, Jae-Yoon Jung, and Min Sim.
\newblock A two-step approach to solar power generation prediction based on weather data using machine learning.
\newblock \emph{Sustainability}, 11\penalty0 (5):\penalty0 1501, 2019.
\newblock ISSN 2071-1050.
\newblock \doi{10.3390/su11051501}.
\newblock URL \url{https://www.mdpi.com/2071-1050/11/5/1501}.

\bibitem[Kraskov et~al.(2004)Kraskov, St\"ogbauer, and Grassberger]{kraskov_2004}
Alexander Kraskov, Harald St\"ogbauer, and Peter Grassberger.
\newblock Estimating mutual information.
\newblock \emph{Phys. Rev. E}, 69:\penalty0 066138, Jun 2004.
\newblock \doi{10.1103/PhysRevE.69.066138}.
\newblock URL \url{https://link.aps.org/doi/10.1103/PhysRevE.69.066138}.

\bibitem[Krechowicz et~al.(2022)Krechowicz, Krechowicz, and Poczeta]{krechowicz_2022}
Adam Krechowicz, Maria Krechowicz, and Katarzyna Poczeta.
\newblock Machine learning approaches to predict electricity production from renewable energy sources.
\newblock \emph{Energies}, 15\penalty0 (23):\penalty0 9146, 2022.
\newblock ISSN 1996-1073.
\newblock \doi{10.3390/en15239146}.
\newblock URL \url{https://www.mdpi.com/1996-1073/15/23/9146}.

\bibitem[Lim et~al.(2022)Lim, Huh, Hong, Park, and Kim]{lim_solar_2022}
Su-Chang Lim, Jun-Ho Huh, Seok-Hoon Hong, Chul-Young Park, and Jong-Chan Kim.
\newblock Solar power forecasting using {CNN}-{LSTM} hybrid model.
\newblock \emph{Energies}, 15\penalty0 (21):\penalty0 8233, 2022.
\newblock ISSN 1996-1073.
\newblock \doi{10.3390/en15218233}.
\newblock URL \url{https://www.mdpi.com/1996-1073/15/21/8233}.

\bibitem[Liu and Chen(2019)]{liu_data_2019}
Hui Liu and Chao Chen.
\newblock Data processing strategies in wind energy forecasting models and applications: A comprehensive review.
\newblock \emph{Applied Energy}, 249:\penalty0 392--408, 2019.
\newblock ISSN 03062619.
\newblock \doi{10.1016/j.apenergy.2019.04.188}.
\newblock URL \url{https://linkinghub.elsevier.com/retrieve/pii/S0306261919308517}.

\bibitem[Liu et~al.(2023{\natexlab{a}})Liu, He, Wu, Liu, Zhang, Chen, Shen, and Li]{laibao_2023}
Laibao Liu, Gang He, Mengxi Wu, Gang Liu, Haoran Zhang, Ying Chen, Jiashu Shen, and Shuangcheng Li.
\newblock Climate change impacts on planned supply–demand match in global wind and solar energy systems.
\newblock \emph{Nature Energy}, 8:\penalty0 1--11, 07 2023{\natexlab{a}}.
\newblock \doi{10.1038/s41560-023-01304-w}.

\bibitem[Liu et~al.(2023{\natexlab{b}})Liu, He, Wu, Liu, Zhang, Chen, Shen, and Li]{liu_climate_2023}
Laibao Liu, Gang He, Mengxi Wu, Gang Liu, Haoran Zhang, Ying Chen, Jiashu Shen, and Shuangcheng Li.
\newblock Climate change impacts on planned supply–demand match in global wind and solar energy systems.
\newblock \emph{Nature Energy}, 8\penalty0 (8):\penalty0 870--880, July 2023{\natexlab{b}}.
\newblock ISSN 2058-7546.
\newblock \doi{10.1038/s41560-023-01304-w}.
\newblock URL \url{https://www.nature.com/articles/s41560-023-01304-w}.

\bibitem[López~Gómez et~al.(2020)López~Gómez, Ogando~Martínez, Troncoso~Pastoriza, Febrero~Garrido, Granada~Álvarez, and Orosa~García]{lopez_gomez_photovoltaic_2020}
Javier López~Gómez, Ana Ogando~Martínez, Francisco Troncoso~Pastoriza, Lara Febrero~Garrido, Enrique Granada~Álvarez, and José~Antonio Orosa~García.
\newblock Photovoltaic power prediction using artificial neural networks and numerical weather data.
\newblock \emph{Sustainability}, 12\penalty0 (24):\penalty0 10295, 2020.
\newblock ISSN 2071-1050.
\newblock \doi{10.3390/su122410295}.
\newblock URL \url{https://www.mdpi.com/2071-1050/12/24/10295}.

\bibitem[Malistov and Trushin(2019)]{malistov_2019}
Alexey Malistov and Arseniy Trushin.
\newblock Gradient boosted trees with extrapolation.
\newblock In \emph{2019 18th IEEE International Conference On Machine Learning And Applications (ICMLA)}, pages 783--789, 2019.
\newblock \doi{10.1109/ICMLA.2019.00138}.

\bibitem[Malvoni et~al.(2016)Malvoni, De~Giorgi, and Congedo]{malvoni_data_2016}
M.~Malvoni, M.G. De~Giorgi, and P.M. Congedo.
\newblock Data on photovoltaic power forecasting models for mediterranean climate.
\newblock \emph{Data in Brief}, 7:\penalty0 1639--1642, 2016.
\newblock ISSN 23523409.
\newblock \doi{10.1016/j.dib.2016.04.063}.
\newblock URL \url{https://linkinghub.elsevier.com/retrieve/pii/S2352340916302773}.

\bibitem[Malvoni et~al.(2017)Malvoni, De~Giorgi, and Congedo]{malvoni_forecasting_2017}
Maria Malvoni, Maria~Grazia De~Giorgi, and Paolo~Maria Congedo.
\newblock Forecasting of {PV} power generation using weather input data‐preprocessing techniques.
\newblock \emph{Energy Procedia}, 126:\penalty0 651--658, 2017.
\newblock ISSN 18766102.
\newblock \doi{10.1016/j.egypro.2017.08.293}.
\newblock URL \url{https://linkinghub.elsevier.com/retrieve/pii/S1876610217338055}.

\bibitem[Mayer and Gróf(2021)]{mayer_2021}
Martin~János Mayer and Gyula Gróf.
\newblock Extensive comparison of physical models for photovoltaic power forecasting.
\newblock \emph{Applied Energy}, 283:\penalty0 116239, 2021.
\newblock ISSN 0306-2619.
\newblock \doi{https://doi.org/10.1016/j.apenergy.2020.116239}.
\newblock URL \url{https://www.sciencedirect.com/science/article/pii/S0306261920316330}.

\bibitem[{Ministère de la Transition Ecologique}(2019)]{ppe}
{Ministère de la Transition Ecologique}.
\newblock Programmations pluriannuelles de l’énergie {(PPE)}.
\newblock \url{https://www.ecologie.gouv.fr/politiques-publiques/programmations-pluriannuelles-lenergie-ppe}, 2019.
\newblock Accessed: August 26, 2024.

\bibitem[{Ministère de la Transition Ecologique}(2020)]{snbc}
{Ministère de la Transition Ecologique}.
\newblock Stratégie nationale bas-carbone {(SNBC)}.
\newblock \url{https://www.ecologie.gouv.fr/politiques-publiques/strategie-nationale-bas-carbone-snbc}, 2020.
\newblock Accessed: August 26, 2024.

\bibitem[Numata and Tanaka(2020)]{numata_2020}
Kohei Numata and Kenichi Tanaka.
\newblock Stochastic threshold model trees: A tree-based ensemble method for dealing with extrapolation.
\newblock \emph{CoRR}, 09 2020.
\newblock \doi{10.48550/arXiv.2009.09171}.

\bibitem[Raymaekers et~al.(2024)Raymaekers, Rousseeuw, Verdonck, and Yao]{raymaekers_2024}
Jakob Raymaekers, Peter Rousseeuw, Tim Verdonck, and Ruicong Yao.
\newblock Fast linear model trees by pilot.
\newblock \emph{Machine Learning}, 113:\penalty0 1--50, 07 2024.
\newblock \doi{10.1007/s10994-024-06590-3}.

\bibitem[Ritchie and Rosado(2020)]{owid-electricity-mix}
Hannah Ritchie and Pablo Rosado.
\newblock Electricity mix.
\newblock \emph{Our World in Data}, 2020.
\newblock https://ourworldindata.org/electricity-mix.

\bibitem[Ryu et~al.(2022)Ryu, Lee, Park, Hwang, Park, Lee, and Kwon]{ryu_evaluation_2022}
Ju-Yeol Ryu, Bora Lee, Sungho Park, Seonghyeon Hwang, Hyemin Park, Changhyeong Lee, and Dohyeon Kwon.
\newblock Evaluation of weather information for short-term wind power forecasting with various types of models.
\newblock \emph{Energies}, 15\penalty0 (24):\penalty0 9403, 2022.
\newblock ISSN 1996-1073.
\newblock \doi{10.3390/en15249403}.
\newblock URL \url{https://www.mdpi.com/1996-1073/15/24/9403}.

\bibitem[Sharma et~al.(2011)Sharma, Sharma, Irwin, and Shenoy]{sharma_predicting_2011}
Navin Sharma, Pranshu Sharma, David Irwin, and Prashant Shenoy.
\newblock Predicting solar generation from weather forecasts using machine learning.
\newblock In \emph{2011 {IEEE} International Conference on Smart Grid Communications ({SmartGridComm})}, pages 528--533. {IEEE}, 2011.
\newblock ISBN 978-1-4577-1702-4 978-1-4577-1704-8.
\newblock \doi{10.1109/SmartGridComm.2011.6102379}.
\newblock URL \url{http://ieeexplore.ieee.org/document/6102379/}.

\bibitem[Sweeney et~al.(2020)Sweeney, Bessa, Browell, and Pinson]{sweeney_2020}
Conor Sweeney, Ricardo~J. Bessa, Jethro Browell, and Pierre Pinson.
\newblock The future of forecasting for renewable energy.
\newblock \emph{WIREs Energy and Environment}, 9\penalty0 (2), 2020.
\newblock \doi{https://doi.org/10.1002/wene.365}.
\newblock URL \url{https://wires.onlinelibrary.wiley.com/doi/abs/10.1002/wene.365}.

\bibitem[Tashman(2000)]{tashman_2000}
Leonard~J. Tashman.
\newblock Out-of-sample tests of forecasting accuracy: an analysis and review.
\newblock \emph{International Journal of Forecasting}, 16\penalty0 (4):\penalty0 437--450, 2000.
\newblock ISSN 0169-2070.
\newblock \doi{https://doi.org/10.1016/S0169-2070(00)00065-0}.
\newblock URL \url{https://www.sciencedirect.com/science/article/pii/S0169207000000650}.
\newblock The M3- Competition.

\bibitem[Taylor(2010)]{taylor_2010}
James~W. Taylor.
\newblock Triple seasonal methods for short-term electricity demand forecasting.
\newblock \emph{European Journal of Operational Research}, 204\penalty0 (1):\penalty0 139--152, 2010.
\newblock ISSN 0377-2217.
\newblock \doi{https://doi.org/10.1016/j.ejor.2009.10.003}.
\newblock URL \url{https://www.sciencedirect.com/science/article/pii/S037722170900705X}.

\bibitem[Teste et~al.(2024)Teste, Makowski, Bazzi, and Ciais]{teste_2024}
Florian Teste, David Makowski, Hassan Bazzi, and Philippe Ciais.
\newblock Early forecasting of corn yield and price variations using satellite vegetation products.
\newblock \emph{Computers and Electronics in Agriculture}, 221:\penalty0 108962, 2024.
\newblock ISSN 0168-1699.
\newblock \doi{https://doi.org/10.1016/j.compag.2024.108962}.
\newblock URL \url{https://www.sciencedirect.com/science/article/pii/S0168169924003533}.

\bibitem[Tsai et~al.(2023)Tsai, Hong, Tu, Lin, and Chen]{tsai_review_2023}
Wen-Chang Tsai, Chih-Ming Hong, Chia-Sheng Tu, Whei-Min Lin, and Chiung-Hsing Chen.
\newblock A review of modern wind power generation forecasting technologies.
\newblock \emph{Sustainability}, 15\penalty0 (14):\penalty0 10757, 2023.
\newblock ISSN 2071-1050.
\newblock \doi{10.3390/su151410757}.
\newblock URL \url{https://www.mdpi.com/2071-1050/15/14/10757}.

\bibitem[{United Nations Convention on Climate Change}(2015)]{paris_agreement}
{United Nations Convention on Climate Change}.
\newblock {Paris Agreement: Climate Change Conference (COP21)}.
\newblock \url{https://unfccc.int/documents/184656}, 2015.
\newblock {Accessed: August 26, 2024}.

\bibitem[Vladislavleva et~al.(2011)Vladislavleva, Friedrich, Neumann, and Wagner]{vladislavleva_predicting_2011}
Katya Vladislavleva, Tobias Friedrich, Frank Neumann, and Markus Wagner.
\newblock Predicting the energy output of wind farms based on weather data: Important variables and their correlation, 2011.
\newblock URL \url{http://arxiv.org/abs/1109.1922}.

\bibitem[Wang et~al.(2019{\natexlab{a}})Wang, Lei, Zhang, Zhou, and Peng]{wang_review_2019}
Huaizhi Wang, Zhenxing Lei, Xian Zhang, Bin Zhou, and Jianchun Peng.
\newblock A review of deep learning for renewable energy forecasting.
\newblock \emph{Energy Conversion and Management}, 198:\penalty0 111799, 2019{\natexlab{a}}.
\newblock ISSN 01968904.
\newblock \doi{10.1016/j.enconman.2019.111799}.
\newblock URL \url{https://linkinghub.elsevier.com/retrieve/pii/S0196890419307812}.

\bibitem[Wang et~al.(2019{\natexlab{b}})Wang, Zhong, Lai, Xia, Wang, and Kang]{wang_exploring_2019}
Jianxiao Wang, Haiwang Zhong, Xiaowen Lai, Qing Xia, Yang Wang, and Chongqing Kang.
\newblock Exploring key weather factors from analytical modeling toward improved solar power forecasting.
\newblock \emph{{IEEE} Trans. Smart Grid}, 10\penalty0 (2):\penalty0 1417--1427, 2019{\natexlab{b}}.
\newblock ISSN 1949-3053, 1949-3061.
\newblock \doi{10.1109/TSG.2017.2766022}.
\newblock URL \url{https://ieeexplore.ieee.org/document/8080240/}.

\bibitem[Wood(2024)]{wood_2024}
Simon~N. Wood.
\newblock On neighbourhood cross validation, 2024.
\newblock URL \url{https://arxiv.org/abs/2404.16490}.

\bibitem[Wood et~al.(2014)Wood, Goude, and Shaw]{wood_2014}
Simon~N. Wood, Yannig Goude, and Simon Shaw.
\newblock {Generalized Additive Models for Large Data Sets}.
\newblock \emph{Journal of the Royal Statistical Society Series C: Applied Statistics}, 64\penalty0 (1):\penalty0 139--155, 05 2014.
\newblock ISSN 0035-9254.
\newblock \doi{10.1111/rssc.12068}.
\newblock URL \url{https://doi.org/10.1111/rssc.12068}.

\bibitem[Yasuda et~al.(2022)Yasuda, Bird, Carlini, Eriksen, Estanqueiro, Flynn, Fraile, {Gómez Lázaro}, Martín-Martínez, Hayashi, Holttinen, Lew, McCam, Menemenlis, Miranda, Orths, Smith, Taibi, and Vrana]{yasuda_2022}
Yoh Yasuda, Lori Bird, Enrico~Maria Carlini, Peter~Børre Eriksen, Ana Estanqueiro, Damian Flynn, Daniel Fraile, Emilio {Gómez Lázaro}, Sergio Martín-Martínez, Daisuke Hayashi, Hannele Holttinen, Debra Lew, John McCam, Nickie Menemenlis, Raul Miranda, Antje Orths, J.~Charles Smith, Emanuele Taibi, and Til~Kristian Vrana.
\newblock C-e (curtailment – energy share) map: An objective and quantitative measure to evaluate wind and solar curtailment.
\newblock \emph{Renewable and Sustainable Energy Reviews}, 160:\penalty0 112212, 2022.
\newblock ISSN 1364-0321.
\newblock \doi{https://doi.org/10.1016/j.rser.2022.112212}.
\newblock URL \url{https://www.sciencedirect.com/science/article/pii/S1364032122001356}.

\bibitem[Zeiler and Fergus(2013)]{zeiler2013}
Matthew~D Zeiler and Rob Fergus.
\newblock Visualizing and understanding convolutional networks, 2013.
\newblock URL \url{https://arxiv.org/abs/1311.2901}.

\bibitem[Zhang et~al.(2019)Zhang, Nettleton, and Zhu]{haozhe_2019}
Haozhe Zhang, Dan Nettleton, and Zhengyuan Zhu.
\newblock Regression-enhanced random forests, 2019.
\newblock URL \url{https://arxiv.org/abs/1904.10416}.

\bibitem[Zhong and Wu(2020)]{zhong_short-term_2020}
You-Jing Zhong and Yuan-Kang Wu.
\newblock Short-term solar power forecasts considering various weather variables.
\newblock In \emph{2020 International Symposium on Computer, Consumer and Control ({IS}3C)}, pages 432--435. {IEEE}, 2020.
\newblock ISBN 978-1-72819-362-5.
\newblock \doi{10.1109/IS3C50286.2020.00117}.
\newblock URL \url{https://ieeexplore.ieee.org/document/9394178/}.

\bibitem[Zhou et~al.(2022)Zhou, Qiu, Feng, and Liu]{zhou_2022}
Huanyu Zhou, Yingning Qiu, Yanhui Feng, and Jing Liu.
\newblock Power prediction of wind turbine in the wake using hybrid physical process and machine learning models.
\newblock \emph{Renewable Energy}, 198:\penalty0 568--586, 2022.
\newblock ISSN 0960-1481.
\newblock \doi{https://doi.org/10.1016/j.renene.2022.08.004}.
\newblock URL \url{https://www.sciencedirect.com/science/article/pii/S0960148122011703}.

\end{thebibliography}

\end{document}